\documentclass[particles, review, accept, pdftex, moreauthors]{Definitions/mdpi}
\firstpage{1}
\pubvolume{1}
\issuenum{1}
\articlenumber{0}
\pubyear{2023}
\copyrightyear{2023}
\datereceived{2023 Jan 09}
\datepublished{}
\hreflink{https://doi.org/} 

\usepackage{wasysym}
\usepackage[export]{adjustbox}

\Title{$\,$\\[-5ex]\hspace*{\fill}{\normalsize{\sf\emph{Preprint no}.\ JLAB-PHY-23-3744, NJU-INP 069/23}}\\[1.25ex]
Nucleon Resonance Electroexcitation Amplitudes and Emergent Hadron Mass}

\TitleCitation{Nucleon Resonance Electroexcitation and Emergent Hadron Mass}

\Author{Daniel S. Carman $^{1,\dagger}$\orcidA{}, Ralf W. Gothe $^{2,\dagger}$\orcidD{}, Victor I. Mokeev $^{1,\dagger}$\orcidC{}, 
and Craig D.\ Roberts $^{3,4,\dagger}$\orcidB{}\,*}

\AuthorNames{Daniel S. Carman, Ralf W. Gothe, Victor I. Mokeev and Craig D.\ Roberts}

\AuthorCitation{Carman, D.S.; Gothe, R.W.; Mokeev, V.I.; and Roberts, C.D.}

\address{%
$^{1}$ \quad Jefferson Laboratory, 12000 Jefferson Ave., Newport News, VA 23606, USA\\
$^{2}$ \quad University of South Carolina, Department of Physics and Astronomy, 712 Main St., Columbia, SC 29208, USA\\
$^{3}$ \quad School of Physics, Nanjing University, Nanjing, Jiangsu 210093, China\\
$^{4}$ \quad Institute for Nonperturbative Physics, Nanjing University, Nanjing, Jiangsu 210093, China \\
}

\corres{Correspondence -- all authors:\\
\qquad carman@jlab.org (D.\,S.\ Carman); gothe@sc.edu (R.\,W.\ Gothe); mokeev@jlab.org (V.\,I.\ Mokeev);\\
\qquad cdroberts@nju.edu.cn (C.\,D.\ Roberts)
}

\firstnote{These authors contributed equally to this work.}

\abstract{
Understanding the strong interaction dynamics that govern the emergence of hadron mass (EHM) represents a challenging open problem in the Standard
Model. In this paper we describe new opportunities for gaining insight into EHM from results on nucleon resonance ($N^\ast$) electroexcitation 
amplitudes ({\it i.e.} $\gamma_vpN^\ast$ electrocouplings) in the mass range up to 1.8\,GeV for virtual photon four-momentum squared ({\it i.e.} 
photon virtualities $Q^2$) up to 7.5\,GeV$^2$ available from exclusive meson electroproduction data acquired during the 6-GeV era of experiments 
at Jefferson Laboratory (JLab). These results, combined with achievements in the use of continuum Schwinger function methods (CSMs), offer new
opportunities for charting the momentum dependence of the dressed quark mass from results on the $Q^2$-evolution of the $\gamma_vpN^\ast$
electrocouplings. This mass function is one of the three pillars of EHM and its behavior expresses influences of the other two, \textit{viz}.\ the
running gluon mass and momentum-dependent effective charge. A successful description of the $\Delta(1232)3/2^+$ and $N(1440)1/2^+$ electrocouplings 
has been achieved using CSMs with, in both cases, common momentum-dependent mass functions for the dressed quarks, for the gluons, and the same 
momentum-dependent strong coupling. The properties of these functions have been inferred from nonperturbative studies of QCD and confirmed, 
\textit{e.g}., in the description of nucleon and pion elastic electromagnetic form factors. Parameter-free CSM predictions for the electrocouplings 
of the $\Delta(1600)3/2^+$ became available in 2019. The experimental results obtained in the first half of 2022 have confirmed the CSM predictions. 
We also discuss prospects for these studies during the 12-GeV era at JLab using the CLAS12 detector, with experiments that are currently in progress, 
and canvass the physics motivation for continued studies in this area with a possible increase of the JLab electron beam energy up to 22\,GeV. Such 
an upgrade would finally enable mapping of the dressed quark mass over the full range of distances ({\it i.e.} quark momenta) where the dominant part 
of hadron mass and $N^\ast$ structure emerge in the transition from the strongly coupled to perturbative QCD regimes.}


\usepackage[mathscr,scaled=1.15]{urwchancal}
\DeclareFontFamily{OT1}{pzc}{}
\DeclareFontShape{OT1}{pzc}{m}{it}%
{<-> s * [1.15] pzcmi7t}{}
\DeclareMathAlphabet{\mathpzc}{OT1}{pzc}{m}{it}

\begin{document}





\section{Introduction}
\label{intro}
Studies of the strong interaction dynamics that govern the generation of hadron ground and excited states in the regime where the running 
coupling of quantum chromodynamics (QCD) is large, {\it i.e.} $\alpha_s/\pi \approx 1$, known as the strong QCD (sQCD) regime, represent a 
crucial challenge in modern hadron physics~\cite{Proceedings:2020fyd}. The rapid growth of $\alpha_s$ in the transition from the perturbative to 
sQCD domains and particularly its saturation, driven by gluon self-interactions, are predicted by CSMs~\cite{Binosi:2016nme, Cui:2019dwv} and
supported by recent experimental results on the Bjorken sum rule~\cite{Deur:2022msf}. These trends suggest that the generation of hadron structure 
in the sQCD regime is defined by emergent degrees of freedom that are related to the partons of QCD's Lagrangian in a non-trivial manner. While 
this evolution with distance is determined by the QCD Lagrangian, it cannot be analyzed by employing perturbative QCD (pQCD) when $\alpha_s/\pi$ 
becomes comparable with unity. The active degrees of freedom seen in hadron structure and their interactions change substantially with distance at 
the scales where the transition from sQCD to pQCD takes place and the structure of hadron ground and excited states emerges. Understanding how the
active degrees of freedom emerge from the QCD Lagrangian and how their interactions evolve with distance requires the development of nonperturbative
methods capable of making predictions, both in the meson and baryon sectors, that can be confronted with empirical results on hadron structure 
extracted using electromagnetic and hadronic probes. 

A decade of rapid progress in the development and application of CSMs in hadron physics~\cite{Eichmann:2016yit, Fischer:2018sdj, Qin:2020rad, 
Roberts:2020udq, Roberts:2020hiw, Roberts:2021xnz, Roberts:2021nhw, Binosi:2022djx, Papavassiliou:2022wrb, Ding:2022ows, Ferreira:2023fva}, 
complemented by advances in and results from lattice QCD (lQCD) \cite{Blum:2014tka, Boyle:2015exm, Boyle:2017jwu, Gao:2017uox, Oliveira:2018lln,
Boucaud:2018xup, Zafeiropoulos:2019flq, Aguilar:2021lke, Aguilar:2021okw, Pinto-Gomez:2022brg}, have delivered numerous predictions for properties 
of mesons and baryons within a common theoretical framework. Studies of hadron structure from data obtained in experiments with electromagnetic 
probes at JLab~\cite{Proceedings:2020fyd, Aznauryan:2012ba, Horn:2016rip, Burkert:2019bhp, Burkert:2022ioj}, MAMI~\cite{A1:2019mrv, Blomberg:2019caf,
Mihovilovic:2016rkr, A1:2016emg, A1:2013fsc, Sparveris:2013ena}, and Babar and Belle~\cite{BaBar:2014omp, Belle-II:2018jsg}, have provided 
experimental results that can be confronted with predictions from the QCD-connected approaches to hadron structure. More results are expected from
experiments in the ongoing 12-GeV era at JLab \cite{Burkert:2018nvj, Proceedings:2020fyd, Barabanov:2020jvn, Accardi:2020swt} and from planned 
research programs at the US electron ion collider (EIC)~\cite{Aguilar:2019teb, Barabanov:2020jvn, AbdulKhalek:2021gbh, Arrington:2021biu}, the
electron ion collider in China (EicC)~\cite{Chen:2020ijn, Anderle:2021wcy}, and experiments with hadronic probes conducted by the AMBER Collaboration 
at CERN~\cite{Quintans:2022utc}.

Studies of exclusive meson electroproduction in the nucleon resonance excitation region using data from 6-GeV-era experiments at JLab have provided 
the first and still only available comprehensive information on the electroexcitation amplitudes ({\it i.e.} $\gamma_vpN^\ast$ electrocouplings) 
of most nucleon excited states in the mass range up to 1.8\,GeV for photon virtualities $Q^2 < 5\,$GeV$^2$ (or $Q^2 < 7.5\,$GeV$^2$ for the
$\Delta(1232)3/2^+$ and $N(1535)1/2^-$)~\cite{Mokeev:2022xfo, Carman:2020qmb, Aznauryan:2011qj, Villano:2009sn, Dalton:2008aa}. 
Analyses of these results have revealed many facets of strong interactions in the sQCD regime seen in the generation of $N^\ast$ states of different 
quantum numbers with different structural features \cite{Barabanov:2020jvn, Anikin:2015ita, Giannini:2015zia, Braun:2016awp, Aznauryan:2018okk,
Qin:2018dqp, Burkert:2019bhp, Ramalho:2019fwf, Lyubovitskij:2020gjz, Raya:2021pyr, Liu:2022nku, Liu:2022ndb}. These results also enable evaluation of
the resonant contributions to inclusive electron scattering observables \cite{Blin:2021twt, HillerBlin:2019jgp, Klimenko:2022Progress}, substantially
expanding the capability to explore both polarized and unpolarized parton distribution functions (PDFs) of the nucleon for fractional parton 
light-front momenta close to unity. Analyses of the results on $\gamma_vpN^\ast$ electrocouplings within CSMs
\cite{Proceedings:2020fyd, Wilson:2011aa, Segovia:2013uga, Mokeev:2015lda, Segovia:2014aza, Segovia:2015hra,  Segovia:2016zyc, Chen:2018nsg, 
Burkert:2019bhp, Burkert:2019opk, Roberts:2019wov, Lu:2019bjs, Carman:2020qmb, Mokeev:2020vab, Raya:2021pyr} have demonstrated a new and promising
potential for elucidation of the sQCD dynamics that are responsible for the generation of $>$98\% of the visible mass in the Universe.

Explaining the emergence of hadron mass represents one of the most challenging, open problems in the Standard Model (SM). The emergent nature of 
hadron mass is made manifest by a comparison between the measured proton and neutron masses and the sum of the current masses of their valence 
quark constituents. Protons and neutrons are bound systems of three light $u$- and $d$-quarks. The sum of the current masses of these quarks, which 
is generated by Higgs couplings into QCD, accounts for less than 2\% of the measured nucleon masses (see Table~\ref{nucleon_mass}). This accounting
clearly indicates that the overwhelmingly dominant component of the nucleon mass is created by mechanisms other than those associated with the Higgs
boson \cite{Roberts:2020udq, Roberts:2020hiw, Roberts:2021xnz, Roberts:2021nhw, Binosi:2022djx, Papavassiliou:2022wrb, Ding:2022ows, Ferreira:2023fva}.

\begin{table*}[t]
\begin{center}
\vspace{2mm}
\begin{tabular}{|c|c|c|} \hline
                    & Proton          & Neutron \\ \hline
Measured masses     & 938.2720813     & 939.5654133 \\
(MeV)               & $\pm$ 0.0000058 & $\pm$ 0.0000058 \\  \hline            
 \begin{tabular}{l}
    Sum of the current \\
    quark masses (MeV)
  \end{tabular} &
  \begin{tabular}{l}
   8.09$^{+1.45}_{-0.65}$ 
  \end{tabular} &
  \begin{tabular}{l}
   11.50$^{+1.45}_{-0.60}$ 
  \end{tabular}
  \\ \hline
Contribution of the current   &          &          \\
quark masses to the measured  &  $<$1.1  & $<$1.4   \\ 
nucleon mass (\%)             &          &          \\ \hline
\end{tabular}
\end{center}
\caption{Comparison between the measured masses of the proton and neutron, $m_{p,n}$, and the sum of the current-quark masses of their three $u$- 
and $d$-quark constituents \cite{ParticleDataGroup:2022pth}.  (Current quark masses listed at a scale of 2\,GeV, but the comparison remains
qualitatively unchanged if renormalization group invariant current masses are used.)
\label{nucleon_mass}}
\end{table*}

The past decade of progress using CSMs to study the evolution of hadron structure with distance, maintaining a traceable and often direct connection 
to the QCD Lagrangian, has conclusively demonstrated that the dominant part of each hadron's mass is generated by strong interactions in the regime 
of large QCD running coupling~\cite{Roberts:2019wov, Proceedings:2020fyd, Roberts:2020udq, Roberts:2020hiw, Roberts:2021xnz, Roberts:2021nhw,
Binosi:2022djx, Papavassiliou:2022wrb, Ding:2022ows, Ferreira:2023fva}. Solving QCD's equations-of-motion for the gluon and quark fields has revealed 
the emergence of quasiparticles, with the quantum numbers of the Lagrangian partons but carrying momentum-dependent masses that are large in the 
sQCD domain. It is the presence of these quasiparticles within mesons and baryons that explains the greatest part of the visible mass in the Universe.

Herein we describe advances in the exploration of the structure of nucleon excited states, using data from the 6-GeV-era experiments at JLab, and 
discuss the impact of these results on the understanding of EHM. A successful description of JLab results on the $\Delta(1232)3/2^+$ and 
$N(1440)1/2^+$ electrocouplings has been achieved using CSMs~\cite{Wilson:2011aa, Segovia:2013uga, Segovia:2014aza, Segovia:2015hra, Chen:2018nsg}. 
The CSM calculations are distinguished by (\textit{a}) having employed common momentum-dependent mass functions for the dressed quarks, whose behavior 
is intimately connected with the running of the gluon mass and QCD's effective charge~\cite{Binosi:2016wcx} and {(\textit{b})} thereby unifying the
description of these electrocouplings with kindred studies of, \textit{inter alia}, nucleon and pion elastic form factors~\cite{Segovia:2014aza,
Chang:2013nia, Cui:2020rmu}. Such successes provide a sound foundation for arguments supporting the potential of experimental results on the 
$Q^2$-dependence of nucleon resonance electrocouplings to deliver new information on the running quark mass. 

Parameter-free CSM predictions for the $Q^2$-evolution of the $\Delta(1600)3/2^+$ electrocouplings became available in 2019~\cite{Lu:2019bjs}. 
At that time, there were no experimental results for the electroexcitation amplitudes of this resonance. Herein, too, we present the first preliminary
experimental results on these amplitudes, obtained from analysis of $\pi^+\pi^-p$ electroproduction data off protons in the $W$-range up to 1.7\,GeV 
and $2<Q^2/{\rm GeV}^2<5$ using the JLab-Moscow State University (JM) meson-baryon reaction model
\cite{Ripani:2000va, Burkert:2007kn, Mokeev:2008iw, CLAS:2012wxw, Mokeev:2015lda}.  Comparison between the CSM predictions and these experimental 
results represents a further sensitive test of the capability for validating the EHM paradigm~\cite{Roberts:2020udq, Roberts:2020hiw, Roberts:2021xnz,
Roberts:2021nhw, Binosi:2022djx, Papavassiliou:2022wrb, Ding:2022ows, Ferreira:2023fva}.

Our discussion is organized as follows. In Section~\ref{csm_basics}, the basic features of the EHM paradigm are outlined, with special emphasis on 
the dressed quark and gluon running masses and their evolution in the transition from the weak to strong coupling domains of the strong interaction. 
We also emphasize the complementarity and critical role of combined studies of both meson and baryon structure in validating any understanding of the
generation of the dressed masses of gluon and quark quasiparticles. We outline the analysis framework used for extraction of the $\gamma_vpN^\ast$ 
electrocouplings from data on exclusive meson electroproduction available from experiments during the 6-GeV era at JLab in 
Section~\ref{res_electrocouplings}. The impact of these results on the understanding of EHM is presented in Section \ref{ehm_resel}. In 
Section~\ref{prospects}, plans for future studies and their prospects in ongoing experiments in the 12-GeV era at JLab are highlighted, along with
physics motivations for a possible increase of the JLab electron beam energy up to 22\,GeV. Such an upgrade would offer the only foreseeable 
opportunities to explore QCD dynamics in the full range of distances over which the dominant part of hadron mass and structure emerge, particularly
reaching, for the first time, into the kinematic region where perturbative and nonperturbative QCD calculations overlap.

\begin{figure}[t]   
\centering
\includegraphics[width=0.75\textwidth, left]{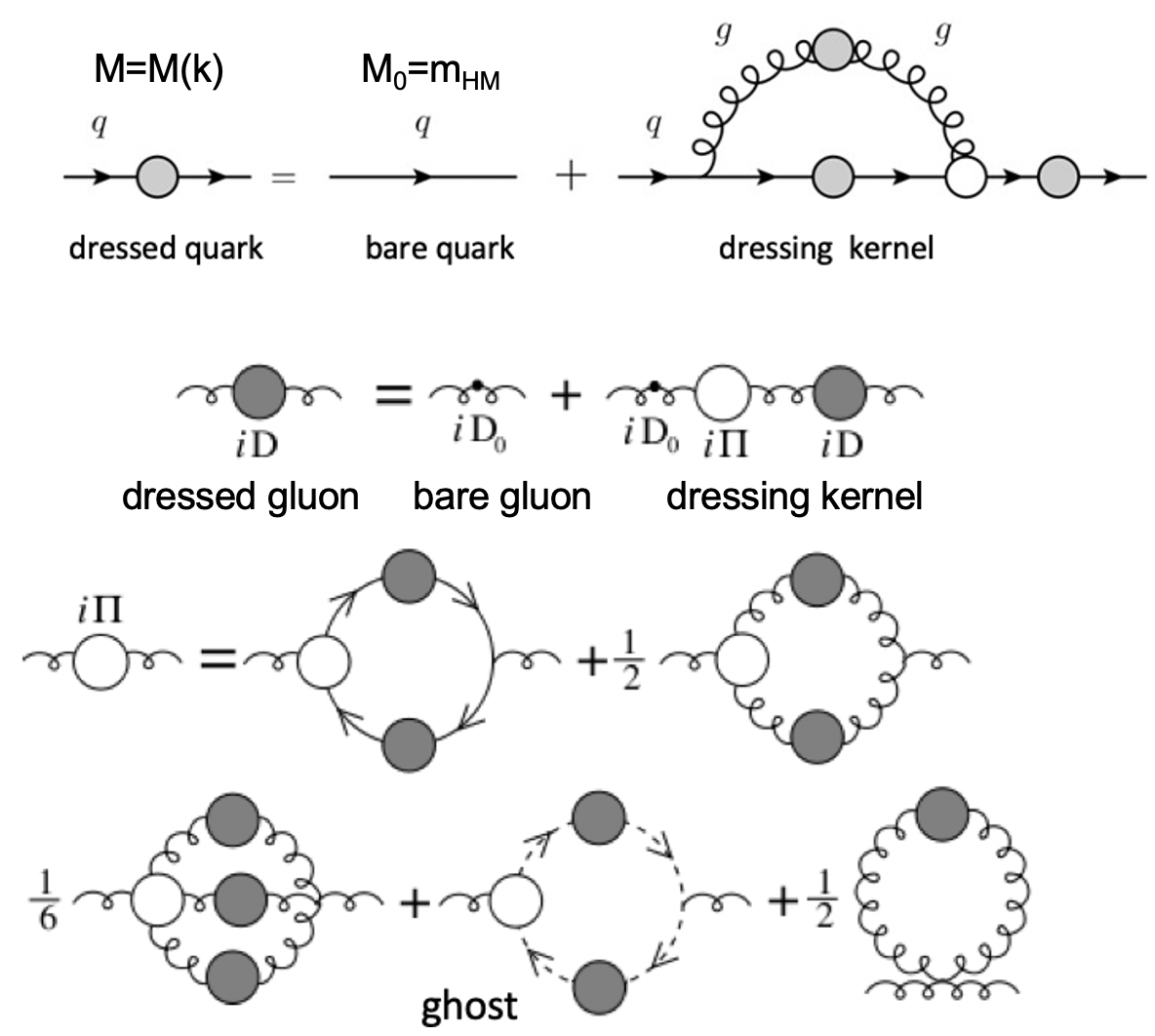}
\caption{\label{gap_equations}
Integral equations for the dressed quark and gluon two-point functions \cite{Roberts:1994dr} (see Section 2.2), drawn in terms of the Feynman 
diagrams that govern the emergence of gluon and quark quasiparticles from the partons used to express the QCD Lagrangian.  These quasiparticles are 
the active components in hadron structure at low resolving scales. Their parton content is revealed at higher resolutions. (Unbroken lines -- quarks;
spring-like lines -- gluons; short-dashed lines -- ghosts; filled circles -- dressed propagators; open circles -- two-point $=$ self-energies and
three/four-point $=$ dressed vertices.  The vertices satisfy their own Dyson-Schwinger equations, involving higher $n$-point functions
\cite{Roberts:1994dr}.)}
\end{figure}

\section{Basics for Insight into EHM Using CSMs}
\label{csm_basics}

The notable progress in developing an understanding of EHM via CSMs has conclusively demonstrated that the dominant part of hadron mass is generated 
by strong interactions at momentum scales $k\lesssim 2\,$GeV. We now sketch the EHM paradigm and discuss how studies of meson and baryon structure
of both ground and excited states offer complementary and crucial information that will enable elucidation of the sQCD dynamics responsible for EHM 
and its manifold corollaries. 

\subsection{CSMs and the EHM Paradigm}
\label{csm_consept}

Every scheme proposed for the solution of QCD reveals that the current-quark masses, which are generated by Higgs boson couplings into the 
Lagrangian, acquire momentum-dependent corrections owing to gluon emission and absorption, as illustrated in Figure~\ref{gap_equations} (top row). 
Gluons, too, come to be dressed by the analogous processes shown in Figure~\ref{gap_equations} (lower rows). Treated in a weak coupling expansion, 
these ``gap equations'' generate every diagram in perturbation theory. On the other hand, nonperturbative analyses can reveal emergent features of 
the strong interaction, such as dynamical chiral symmetry breaking (DCSB) and intimations of confinement \cite{Ding:2022ows} (Section 5).

\begin{figure}[t]
\includegraphics[width=0.85\textwidth, left]{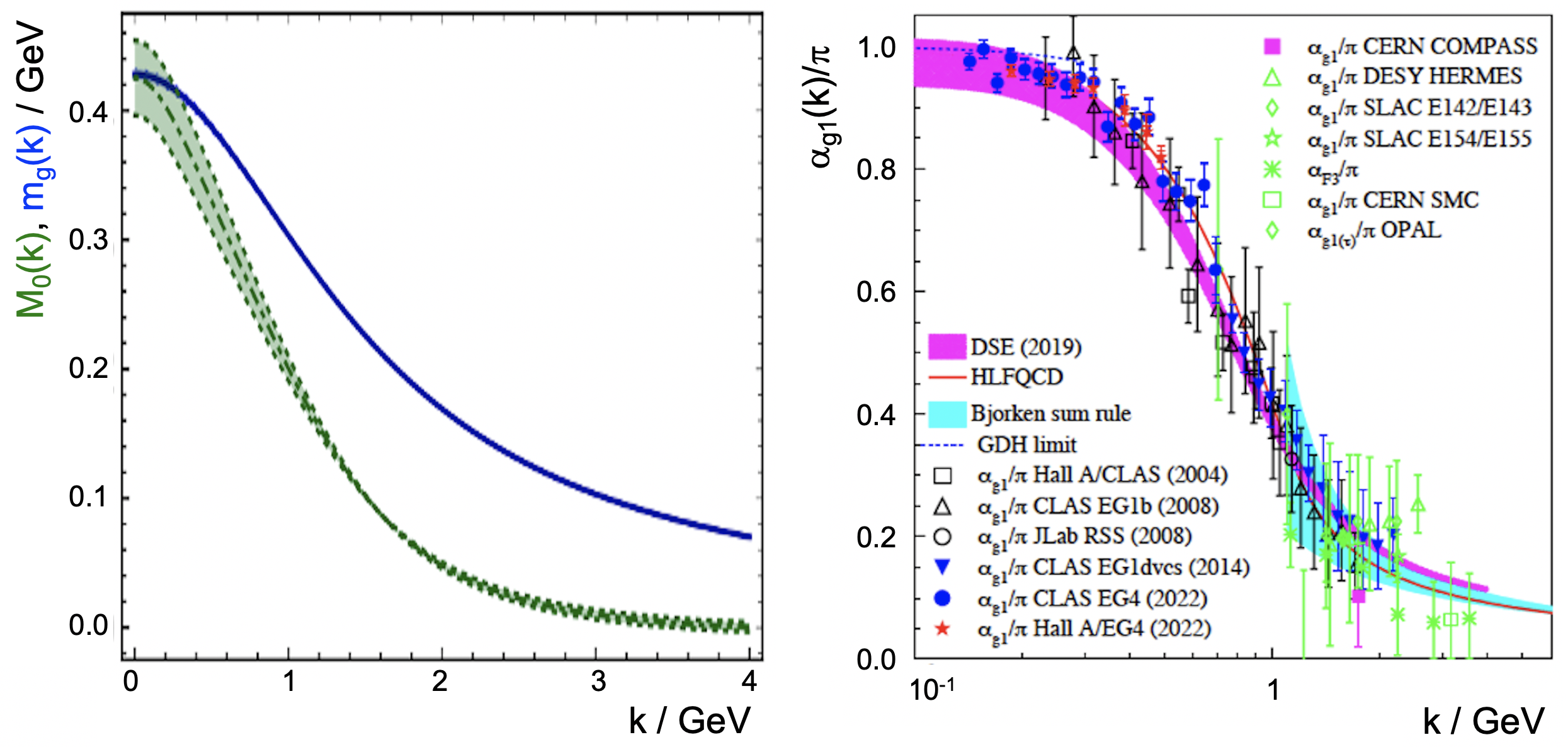}
\caption{\label{alpha_mass_running}
(Left): CSM predictions for the momentum dependence of the dressed gluon (blue solid) and quark (green dot-dashed) masses
\cite{Roberts:2021xnz,Roberts:2020hiw, Roberts:2021nhw}. The associated like-colored bands express the uncertainties in the CSM predictions. 
(\textit{N.B}.\ Since the Poincar\'e-invariant kinetic energy operator for a vector boson has mass-dimension two and that for a spin-half 
fermion has mass-dimension unity, then for $m_p^2/k^2 \to 0$, $M_0(k) \propto 1/k^2$ and $m_g^2(k) \propto 1/k^2$, up to $\ln k^2$ corrections).
(Right) CSM prediction~\cite{Cui:2019dwv} (magenta band) for the process-independent QCD running coupling $\hat\alpha_s(Q)$ compared with the
empirical results~\cite{Deur:2022msf} for the process-dependent effective charge defined via the Bjorken sum rule, which is prominent in deep 
inelastic scattering.}
\end{figure}

As explained elsewhere~\cite{Roberts:1994dr}, the integral equations in Figure~\ref{gap_equations}, and their analogs for higher-$n$-point 
functions, can be understood as QCD's Euler-Lagrange equations, \textit{viz}.\ QCD's equations of motion. The solutions of those shown explicitly 
predict the emergence of gluon and quark quasiparticles, each of which is a superposition of enumerably many gluon and quark (and ghost) partons, is
characterized by its own momentum-dependent mass function -- see the left panel of Figure~\ref{alpha_mass_running}, and evolves with distance $1/k$,
where $k$ is the momentum-scale flowing through the diagram, in a well-defined manner that reproduces perturbative results on $m_p/k\simeq 0$.

Of primary significance is the dressing of gluons, described by the lower three rows in Figure~\ref{gap_equations}, with effects driven by the 
three-gluon vertex being most prominent. It was realized long ago~\cite{Cornwall:1981zr} that this leads to the emergence of a running gluon mass, 
like that in Figure~\ref{alpha_mass_running} (left panel), through the agency of a Schwinger mechanism~\cite{Schwinger:1962tn, Schwinger:1962tp} in 
QCD, the details of which have steadily been unfolded during the past fifteen years~\cite{Aguilar:2008xm, Boucaud:2011ug, Aguilar:2015bud, 
Aguilar:2021uwa, Aguilar:2022thg}. This essentially nonperturbative consequence of gauge sector dynamics, revealed in both continuum and 
lattice-regularized studies of QCD, is the first pillar of EHM.

Capitalizing on such progress in understanding gauge sector dynamics, a unique QCD analog of the Gell-Mann--Low effective charge has been defined 
and calculated~\cite{Binosi:2016nme, Cui:2019dwv}, $\hat\alpha(k)$, with the result shown in Figure~\ref{alpha_mass_running} (right panel). For 
$k\gtrsim 2\,$GeV, this charge matches the pQCD coupling, but it also supplies an infrared completion of the running coupling, which is free of a 
Landau pole and saturates to the value $\hat\alpha(k=0)=0.97(4)$. Both these latter features are direct consequences of the emergence of a gluon 
mass function, whose infrared value is characterized by the renormalization-group-invariant mass-scale $\hat m = 0.43(1)\,$GeV$\approx m_p/2$.
This effective charge is the second pillar of EHM.  

As highlighted in Figure~\ref{alpha_mass_running} (right panel), the pointwise behavior of $\hat\alpha(k)$ is almost identical to that of the 
process-dependent charge~\cite{Grunberg:1982fw, Grunberg:1989xf} defined via the Bjorken sum rule~\cite{Bjorken:1966jh, Bjorken:1969mm} for reasons 
that are explained in Reference~\cite{Ding:2022ows} (Section 4). The form of $\hat\alpha(k)$ -- in particular, its being defined and smooth on 
the entire domain of spacelike momentum transfers -- provides strong support for the conjecture that QCD is a mathematically well-defined quantum 
gauge field theory. As such, it can serve as a template for extensions of the SM using the notion of compositeness for seemingly pointlike objects. 

Turning to the quark gap equation, Figure~\ref{gap_equations} (top row), and constructing its kernel using the first two pillars of EHM, one obtains 
a dressed-quark propagator that is characterized by the mass-function shown in Figure~\ref{alpha_mass_running} (left panel). Critically, for 
$k\lesssim 2\,$GeV, the behavior of this mass function is practically unchanged in the absence of Higgs boson couplings into QCD, \textit{i.e}., in 
the chiral limit. Such an outcome is impossible in pQCD. The emergence of $M(k)$ is the principal manifestation of DCSB in QCD; and this 
dressed-quark mass function, again a prediction common to both continuum and lattice-regularized QCD, is the third pillar of EHM.
 
The appearance of dressed-gluon and -quark quasiparticles following the transition into the domain of sQCD, whose strong mutual- and self-interactions
are described by a process-independent momentum-dependent effective charge, form the basis for the EHM paradigm and its explanation of hadron mass and
structure. Indeed, it is worth reiterating that the dressed-quark mass function in Figure~\ref{alpha_mass_running} (left panel) shows how the almost 
massless current-quark partons, which are the degrees-of-freedom best suited for the description of truly high-energy phenomena, are transmogrified, 
by a nonperturbative accumulation of interactions, into fully dressed quarks. It is these quark quasiparticles, to which is attached an infrared 
mass-scale $M(k \simeq 0) \approx 0.4\,$GeV, that provide a link between QCD and the long line of quark potential models developed in the past sixty
years~\cite{GellMann:1964nj}.

Insofar as the light $u$- and $d$-quarks are concerned, Higgs boson couplings into QCD are almost entirely irrelevant to the size of their infrared 
mass, contributing $< 2$\% -- Reference~\cite{Roberts:2021nhw} (Figure 2.5); hence, equally irrelevant to the masses of the nucleon and its excited 
states. The dominant component of the masses of all light-quark hadrons is that deriving from $M(k\simeq 0)$, \textit{viz}.\ EHM.

Since the quark quasiparticles carry the same quantum numbers as the seed quark-partons, then $N^\ast$ electroexcitation processes can be used to 
chart $M(k)$ by exploiting the dependence of the associated $N^\ast$ electroexcitation amplitudes on the momentum transfer squared. Sketched simply,
owing to the quasielastic nature of the transition, the momentum transferred in the process, $Q$, is shared equally between the three bound quark
quasiparticles in the initial and, subsequently, the final states. This means that the quark mass function is predominantly sampled as $M(Q/3)$, 
because bound-state wave functions are peaked at zero relative momentum. Hence, increasing $Q$ takes the reaction cross section smoothly from the 
sQCD (constituent-quark) domain into the pQCD domain. Any Poincar\'e-invariant, QCD-connected calculational framework can then relate the 
$Q^2$-dependence of the electroexcitation amplitudes to the momentum dependence of the quark mass function and, crucially, when it comes to 
predictions, vice versa. Examples are provided in References~\cite{Wilson:2011aa, Segovia:2013uga, Segovia:2014aza, Segovia:2015hra, Segovia:2016zyc, 
Chen:2018nsg, Eichmann:2018ytt, Burkert:2019bhp, Roberts:2019wov, Lu:2019bjs, Raya:2021pyr}. Regarding $M(k)$ in Figure~\ref{alpha_mass_running}, one
enters the perturbative domain for $k \gtrsim 2\,$GeV, hence, a comprehensive mapping of the nonperturbative part of the dressed-quark mass requires 
\begin{equation}
0\leq Q^2/{\rm GeV}^2 \lesssim 20 - 30\,.
\label{ChartMk}
\end{equation}
Experiments at JLab during the 6-GeV era provided a beginning, their progeny during the 12-GeV era will extend the map further, but only an 
upgrade of the JLab accelerator energy to beyond 20\,GeV will deliver near exhaustive coverage of the full EHM domain.

\subsection{Some Highlights from the EHM Experiment-Theory Connection}
\label{qmass_hadr_struct}

\begin{figure}[t]   
\centering
\includegraphics[width=0.9\textwidth, left]{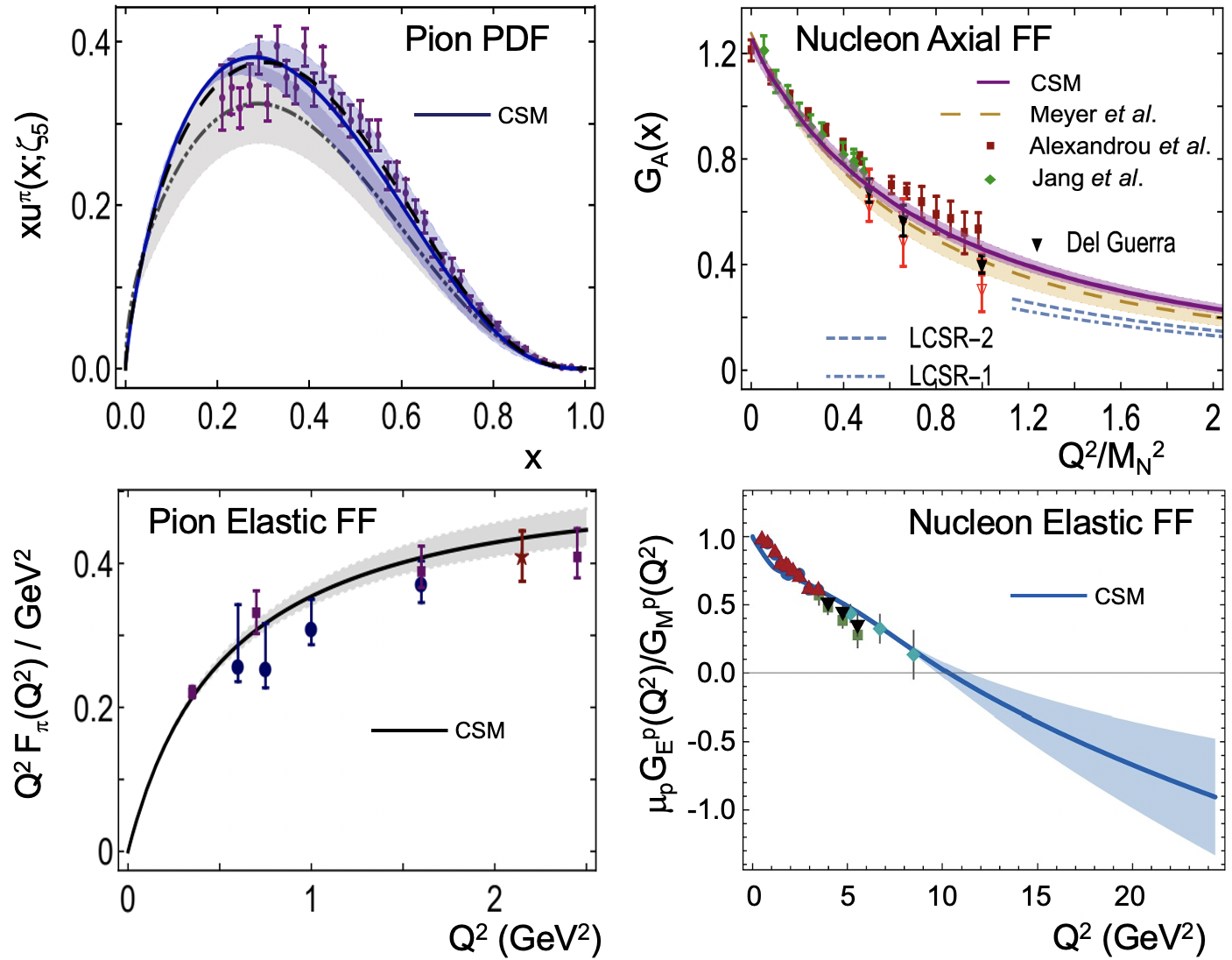}
\caption{\label{csm_predictions}
CSM predictions for observables of the structure for the ground state hadrons in comparison with experimental results (points with error bars). 
\textit{Upper left} -- pion valence quark PDF \cite{Ding:2019lwe}; 
\textit{upper right} -- nucleon axial form factor $G_{A}$ ~\cite{Chen:2022odn};
\textit{lower-left}  -- pion elastic form factor \cite{Roberts:2021nhw}; and
\textit{lower right} -- ratio of nucleon elastic electric and magnetic form factors \cite{Cui:2020rmu}.}
\end{figure}

Charting the dressed-quark mass function using results from hadron structure experiments is a principal goal of modern hadron physics. As always, 
there are challenges to be overcome, but the potential rewards are great. Empirical verification of the EHM paradigm will pave the way to 
understanding the origin of the vast bulk of visible mass in the Universe. As illustration, we note that CSMs have supplied a large body of results 
for meson and baryon structure observables; some examples are shown in Figure~\ref{csm_predictions}. Each of these predictions was obtained within 
a common theoretical framework and expresses different observable consequences of the dressed-gluon and -quark mass functions shown in 
Figure~\ref{alpha_mass_running}, hence, draws a clear connection between observation and the QCD Lagrangian. Notably, here we have only highlighted
results for ground-state hadrons because CSM predictions for the $\gamma_vpN^\ast$ electrocouplings and their comparison with experimental results 
are discussed below. 

Owing to the pattern of DCSB in QCD, a quark-level Goldberger-Treiman identity \cite{Maris:1997hd, Maris:1997tm, Holl:2004fr, Brodsky:2012ku, 
Qin:2014vya}:
\begin{equation}
\label{gt_rel}
f_{\rm NG} E_{\rm NG}(k^2)=B(k^2)\,,
\end{equation}
relates the leading term in the bound-state amplitude of all Nambu-Goldstone (NG) bosons, $E_{\rm NG}(k^2)$, to the scalar piece of the dressed-quark
self energy, $B(k^2)$, with the NG boson leptonic decay constant, $f_{\rm NG}$, providing the constant of proportionality. This exact relationship in
chiral-limit QCD is Poincar\'e-invariant, gauge-covariant, and renormalization-scheme independent. It is also the SM's most fundamental expression of
the Nambu-Goldstone theorem~\cite{Nambu:1960tm, Goldstone:1961eq}. Equation~\ref{gt_rel} explains the seeming dichotomy of massless NG bosons being 
composites built from massive quark and antiquark quasiparticles; ensuring that all one-body dressing effects that give rise to the quasiparticle
masses are cancelled exactly by binding energy with the bound-states, so that they emerge as massless composite objects in the chiral limit
\cite{Roberts:2016vyn}.

Equation~\ref{gt_rel} expresses other remarkable facts. It is also a precise statement of equivalence between the pseudoscalar-meson two-body and 
matter-sector one-body problems in chiral-QCD. These problems are usually considered to be essentially independent. Moreover, it reveals that the 
cleanest expressions of EHM in the SM are located in the properties of the massless NG bosons. It is worth stressing here that $\pi$- and $K$-mesons
are indistinguishable in the absence of Higgs couplings into QCD. Furthermore, as noted above, Equation \ref{gt_rel} entails that they are entirely 
massless in this limit: the $\pi$ and $K$ mesons are the NG bosons that emerge as a consequence of DCSB. At realistic Higgs couplings, however, $\pi$
and $K$ observables are windows onto both EHM and its modulation by Higgs boson couplings into QCD. 

It is now widely recognized~\cite{Horn:2016rip, Aguilar:2019teb, Chen:2020ijn, Anderle:2021wcy, Arrington:2021biu, Quintans:2022utc} that the 
quark-level Goldberger-Treiman identity, Equation \ref{gt_rel}, and its corollaries lift studies of $\pi$ and $K$ structure to the highest level 
of importance. CSM calculations are available for a broad range of such observables; \textit{e.g}., in a challenge for future high-luminosity, 
high-energy facilities, a prediction for the elastic electromagnetic pion form factor is now available out to $Q^2=40\,$GeV$^2$ -- see
Reference~\cite{Aguilar:2019teb} (Figure 9).  

\begin{figure}[t]
\centering
\includegraphics[width=0.7\textwidth, left]{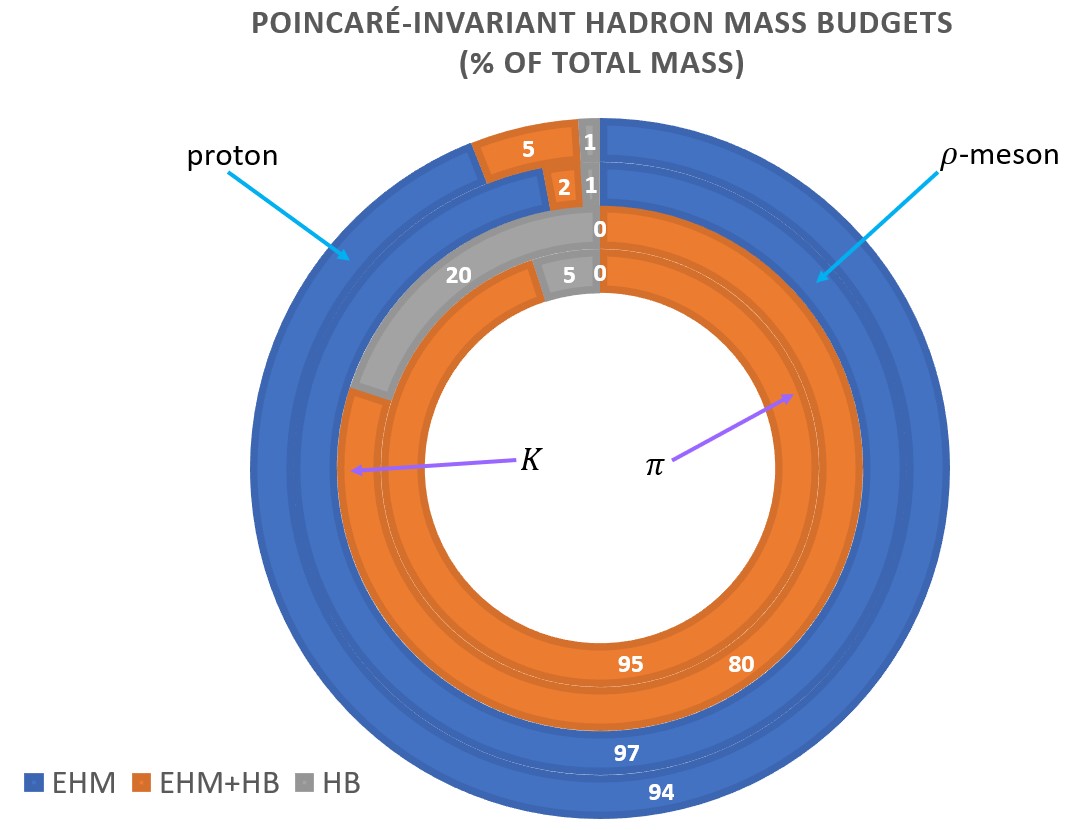}
\caption{\label{Fmassbudget}
Mass budgets for the proton (outermost annulus), $\rho$-meson, kaon, and pion (innermost annulus). Each annulus is drawn using a Poincar\'e-invariant
decomposition. The separation is made at a renormalization scale $\zeta=2\,$GeV, calculated using information from 
References~\cite{Flambaum:2005kc, RuizdeElvira:2017stg, Aoki:2019cca, Workman:2022ynf}.}
\end{figure}

The peculiar character of NG bosons is further highlighted by the mass budgets drawn in Figure~\ref{Fmassbudget}, which identify that component of 
the given hadron's mass that is generated by (\textit{i}) EHM; (\textit{ii}) constructive interference between EHM and the Higgs-boson (HB) mass
contribution; and (\textit{iii}) that part generated solely by the Higgs. The proton annulus depicts information already presented in 
Table~\ref{nucleon_mass} and highlights again that the proton mass owes almost entirely to the mechanisms of EHM. New information is expressed in 
the second annulus, which is the $\rho$-meson mass budget. Plainly, the $\rho$-meson and proton mass budgets are qualitatively and semi-quantitatively 
identical, despite one being a meson and the other a baryon.

The $\pi$ and $K$ mass budgets in Figure~\ref{Fmassbudget} are completely different. For these (near) NG bosons, there is no pure EHM component -- 
no blue part of the ring -- because they are massless in the chiral limit. On the other hand, the HB contribution to the pion mass is commensurate 
with the kindred component of the proton and $\rho$-meson masses. The biggest contribution for the $\pi$ is EHM+HB interference: the small HB-only
contribution is magnified by a huge, latent EHM component. The $K$-meson mass budget is similar. However, the larger current-mass of the $s$-quark 
entails that the HB-alone contribution is four-times larger in the $K$ than in the $\pi$, but it is not $\sim 15$-times larger, as a simple counting 
of current-masses would suggest. Evidently, there is some subtlety in EHM+HB interference effects.

This discussion summarizes what others have explained in detail~\cite{Roberts:2020udq, Roberts:2020hiw, Roberts:2021xnz, Roberts:2021nhw, 
Binosi:2022djx, Papavassiliou:2022wrb, Ding:2022ows, Ferreira:2023fva}, namely, that studies of NG bosons on one hand and the nucleon and its 
excited states on the other provide complementary information about the mechanisms behind EHM: NG bosons reveal much about EHM+HB interference, 
whereas the other systems are directly and especially sensitive to EHM-only effects. It follows that consistent results on the dressed-quark mass
function and, therefrom, indirectly, on the gluon mass and QCD effective coupling, obtained from experimental studies of these complementary systems 
-- NG bosons and the nucleon and its excitations - will shine the brightest light on the many facets and expressions of emergent hadron mass and
structure in Nature. Such a broad approach is the best (only?) way to properly verify the EHM paradigm.

\section{Nucleon Resonance Electrocouplings and their Impact on the Insight into EHM}
\label{res_electrocouplings}

The contemporary application of CSMs provides a QCD-connected framework that enables development of an understanding of EHM~\cite{Roberts:2020udq,
Roberts:2020hiw, Roberts:2021xnz, Roberts:2021nhw, Binosi:2022djx, Papavassiliou:2022wrb, Ding:2022ows, Ferreira:2023fva} from the comparison of 
theory predictions with experimental results on the $Q^2$-evolution of nucleon elastic form factors and nucleon resonance electroexcitation 
amplitudes~\cite{Mokeev:2015lda, Burkert:2019opk, Carman:2020qmb, Mokeev:2020vab}. In this Section, we provide an overview of experimental 
$\gamma_vpN^\ast$ results where comparisons with CSM predictions exist.

\subsection{Extraction of Electrocouplings from Exclusive Meson Electroproduction Data}
\label{elcoupl_exctraction}

Nucleon resonance electroexcitations can be fully described in terms of three electroexcitation amplitudes or $\gamma_vpN^\ast$ electrocouplings. 
$A_{1/2}(Q^2)$ and $A_{3/2}(Q^2)$ describe resonance production in the process $\gamma_v p \to N^\ast, \Delta^\ast$ by transversely polarized 
photons of helicity $+1$ {($-1$)} and target proton helicities $\pm 1/2$ {($\mp 1/2$)} in the center-of-mass (CM) frame, with the resonance spin
projection, directed parallel (antiparallel) to the $\gamma_v$ momentum, equal to $1/2$ ($-1/2$) and $3/2$ ($+3/2$), respectively. The 
resonance electroexcitation amplitudes of the other (flipped) helicities of the initial photon and target proton and the resonance spin projections 
are related by parity transformations. $S_{1/2}(Q^2)$ describes accordingly the resonance electroexcitation by a longitudinal virtual photon of zero
helicity and target proton helicities $\pm 1/2$, with the absolute value of the resonance spin projection equal to 1/2 \cite{Aznauryan:2011qj}.
Since parity is conserved in both electromagnetic and strong interactions, the $A_{1/2}(Q^2)$, $A_{3/2}(Q^2)$, and $S_{1/2}(Q^2)$ electrocouplings
describe all possible $N^\ast$ electroexcitation amplitudes. These electrocouplings are unambiguously determined through their relation with the 
resonance electromagnetic decay widths, $\Gamma_\gamma^T$ and $\Gamma_\gamma^L$, to the final state for transversely and longitudinally polarized
photons:

\begin{subequations}
\begin{align}
\Gamma_\gamma^T(W=M_r,Q^2) & =\frac{q^2_{\gamma,r}(Q^2)}{\pi}\frac{2M_N}{(2J_r+1)M_r} 
\left(|A_{1/2}(Q^2)|^2+|A_{3/2}(Q^2)|^2\right) , \\
\Gamma_\gamma^L(W=M_r,Q^2) & =\frac{q^2_{\gamma,r}(Q^2)}{\pi}\frac{2M_N}{(2J_r+1)M_r}|S_{1/2}(Q^2)|^2,\label{Eq:EMWidths}
\end{align}
\end{subequations}
with $q_{\gamma,r}=\left.q_{\gamma} \right|_{W=M_r}$, the absolute value of the $\gamma_v$ three momentum at the resonance point, $M_r$ and $J_r$ 
being the resonance mass and spin, respectively, and $M_N$ the nucleon mass. $W$ is the sum of the energies of the $\gamma_v$ and target proton in
their CM frame. 

Alternatively, the resonance electroexcitation can be described by three transition form factors, $G_{1,2,3}(Q^2)$ or $G^\ast_{M,E,C}(Q^2)$,  
which represent Lorentz invariant functions in the most general expressions for the $N \to N^\ast$ electromagnetic transition currents. For 
spin 1/2 resonances, the $F^\ast_{1,2}(Q^2)$ Dirac and Pauli transition form factors can also be used instead of $G_{1,2,3}(Q^2)$ or
$G^\ast_{M,E,C}(Q^2)$. The description of resonance electroexcitation in terms of the electrocouplings and the electromagnetic transition form 
factors is completely equivalent, since they are unambiguously related, as described in References~\cite{Aznauryan:2011qj, Obukhovsky:2019xrs}.

\begin{figure}[t]
\includegraphics[width=0.85\textwidth, left]{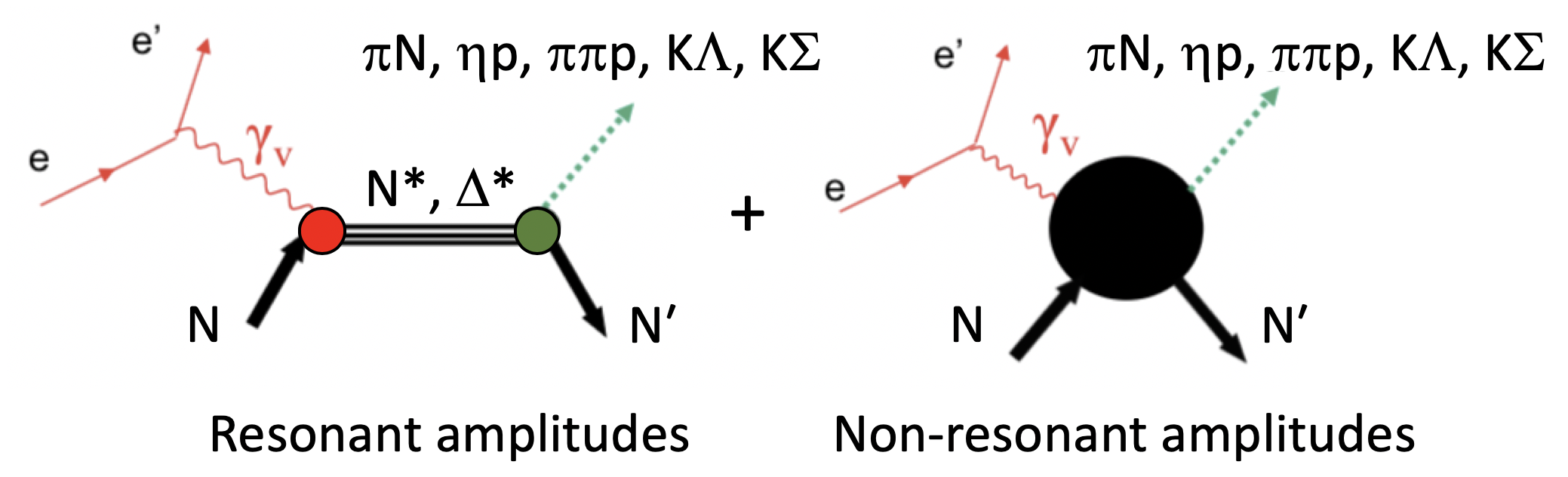}
\vspace{-4mm}
\caption{Resonant and non-resonant amplitudes contributing to exclusive meson electroproduction channels in the resonance region.}
\label{exclusive_reactions1}
\end{figure}

The $\gamma_vpN^\ast$ electrocouplings have been determined from data on exclusive meson electroproduction for most relevant channels in the 
resonance excitation region, including $\pi N$, $\eta p$, and $\pi^+\pi^-p$. The extractions for $KY$ channels are still awaiting the development 
of a reaction model capable of describing electroproduction observables with accuracy sufficient for the reliable separation of the
resonant/non-resonant contributions~\cite{Carman:2016hlp, Carman:2018fsn}. The full amplitude for any exclusive electroproduction channel can be 
described as the coherent sum of the $N^\ast$ electroexcitations in the $s$-channel for the virtual photon-proton interaction and a complex set of 
non-resonant mechanisms, as depicted in Figure~\ref{exclusive_reactions1}. The electrocouplings determined from all exclusive meson 
electroproduction channels should be the same for a given $N^\ast$ state, since they should be independent of their hadronic decays, while the 
non-resonant amplitudes are different for each exclusive meson electroproduction channel. Hence consistent results on the $Q^2$-evolution of the 
electrocouplings extracted from different decay channels enable evaluation of the systematic uncertainties related to the use of the reaction 
models employed in the analysis.

Systematic studies of $N^\ast$ electroexcitation from the data became feasible only after experiments during the 6-GeV era with the CLAS detector 
in Hall~B at JLab. This detector has collected the dominant part of available world data on most single- and multi-meson electroproduction channels 
off protons in the resonance region for $Q^2$ up to 5\,GeV$^2$~\cite{Aznauryan:2011qj,Burkert:2019opk, Mokeev:2020vab, Carman:2020qmb, Mokeev:2022xfo}.
The data are stored in the CLAS Physics Database~\cite{CLAS:DB,Chesnokov:2022gjb}. For the first time, a large body of data ($\approx 150$k points) 
on differential cross sections and polarization asymmetries has become available with nearly complete coverage for the final state hadron CM 
emission angle, which is important for the reliable extraction of electrocouplings.

Several reaction models have been developed for the extraction of electrocouplings from independent studies of the $\pi N$~\cite{Aznauryan:2002gd, 
Aznauryan:2005tp, Arndt:2007yrz, Drechsel:2007if, CLAS:2009ces, Tiator:2011pw, CLAS:2014fml, Tiator:2016btt, Tiator:2017cde, Tiator:2018pjq}, 
$\eta N$ \cite{Knochlein:1995qz, Chiang:2001as, Chiang:2002np, Aznauryan:2003zg, Tiator:2018pjq}, and $\pi^+\pi^-p$~\cite{Ripani:2000va, 
Burkert:2007kn, Mokeev:2008iw, CLAS:2012wxw, Mokeev:2015lda} electroproduction channels off protons. Coupled-channel approaches
\cite{Kamano:2018sfb, Mai:2021aui, Mai:2021vsw} are making steady progress toward determining the electrocouplings from global multichannel analyses 
of the combined data for exclusive meson photo-, electro-, and hadroproduction. These analyses will allow for the explicit incorporation of final 
state interactions between all open channels for the strong interactions between the final state hadrons. Application of such advanced 
coupled-channel approaches will also enable the restrictions imposed on the photo-, electro-, and hadroproduction amplitudes by the general 
unitarity condition to be consistently taken into account. An important extension of the database on the exclusive meson hadroproduction channels 
is expected from the JPARC experimental program~\cite{J-PARCE45:2015azl,JPARC}. These data will be of particular importance in extending the 
extraction of the electrocouplings within global multichannel analyses toward $W > 1.6\,$GeV.

\begin{table}[t]
\begin{center}
\vspace{2mm}
\begin{tabular}{|c|c|c|} \hline
Meson                   & Excited proton  & $Q^2$ ranges for extracted \\
electroproduction       & states          & $\gamma_vpN^\ast$  \\
channels                &                 & electrocouplings, GeV$^2$ \\ \hline                    
$\pi^0 p$, $\pi^+ n$    & $\Delta(1232)3/2^+$    & 0.16-6 \\
                        & $N(1440)1/2^+$, $N(1520)3/2^-$  & 0.30-4.16   \\
                        & $N(1535)1/2^-$  &   0.30-4.16            \\ \hline            
$\pi^+ n$               & $N(1675)5/2^-$, $N(1680)5/2^+$    & 1.6-4.5 \\
                        & $N(1710)1/2^+$               &             \\ \hline
$\eta p$                & $N(1535)1/2^-$        &    0.2-2.9 \\ \hline               
$\pi^+\pi^-p$           & $N(1440)1/2^+$, $N(1520)3/2^-$    & 0.25-1.50 \\
                        & $\Delta(1600)3/2^+$, $\Delta(1620)1/2^-$ & 2.0-5.0   \\
                        & $N(1650)1/2^-$, $N(1680)5/2^+$, & \\ 
                        & $\Delta(1700)3/2^-$  &   0.50-1.50 \\ 
                        & $N(1720)3/2^+$, $N'(1720)3/2^+$  &    0.50-1.50  \\ \hline
\end{tabular}
\end{center}
\caption{
\label{el_clas}Summary of the results for the $\gamma_vpN^\ast$ electrocouplings from the $\pi N$, $\eta p$, and $\pi^+\pi^-p$ electroproduction 
channels measured with the CLAS detector in Hall~B at JLab.}
\end{table}

\begin{figure}[t]
\includegraphics[width=0.85\textwidth, left]{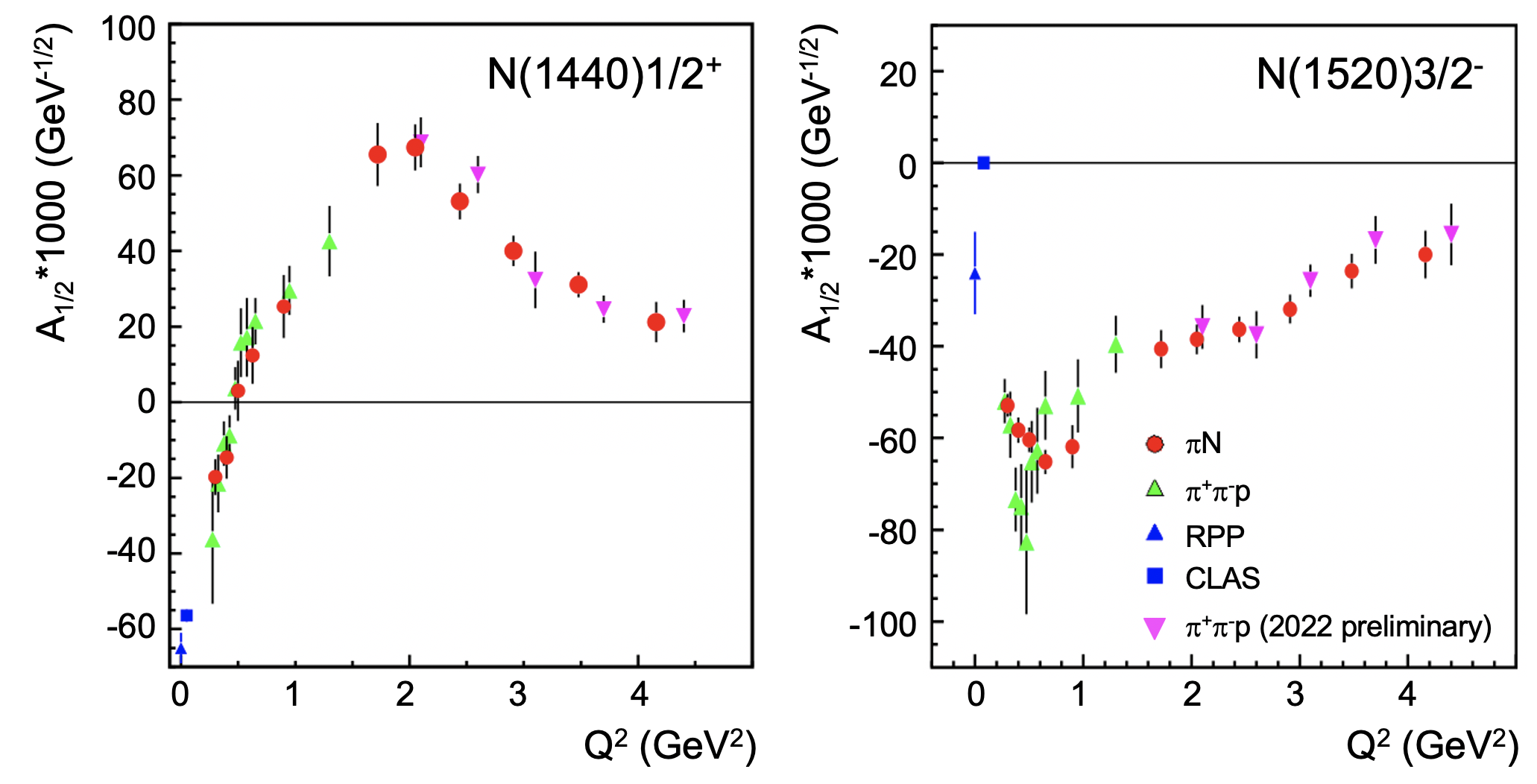}
\vspace{-5mm}
\caption{$N(1440)1/2^+$ and $N(1520)3/2^-$ electrocouplings extracted from the $\pi N$~\cite{Aznauryan:2009mx, CLAS:2014fml} and $\pi^+\pi^-p$ 
\cite{Mokeev:2015lda, CLAS:2012wxw, VictorMokeevPresentation1, VictorMokeevPresentation2} electroproduction channels. The photocouplings from 
the Review of Particle Properties (RPP)~\cite{Workman:2022ynf} and from Reference~\cite{Dugger:2009pn} are shown by the blue squares and triangles,
respectively.}
\label{p11d13_elcoupl}
\end{figure}

Analyses of CLAS results from the exclusive $\pi N$, $\eta p$, and $\pi^+\pi^-p$ electroproduction channels have provided the first and still 
only available comprehensive information on the electrocouplings of most excited proton states in the range of $W < 1.8$~GeV and $Q^2 < 5$~GeV$^2$ 
(see Table~\ref{el_clas}). The experiments of the 6-GeV era in Halls~A/C at JLab further extended this information, providing $\Delta(1232)3/2^+$ 
and $N(1535)1/2^-$ electrocouplings for $Q^2 < 7$~GeV$^2$~\cite{Villano:2009sn,Dalton:2008aa}.

As representative examples, the transverse $A_{1/2}(Q^2)$ electrocouplings versus $Q^2$ for the $N(1440)1/2^+$ and $N(1520)3/2^-$, obtained from
independent studies of the $\pi N$~\cite{Aznauryan:2009mx, CLAS:2014fml} and $\pi^+\pi^-p$~\cite{Mokeev:2015lda,CLAS:2012wxw} channels, are shown 
in Figure~\ref{p11d13_elcoupl}. The electrocouplings inferred from data on the two major $\pi N$ and $\pi^+\pi^-p$ electroproduction channels, 
with different non-resonant contributions, are consistent. This success, reproduced for all available electrocouplings and reaction channels (see 
Table \ref{el_clas}), has demonstrated the capabilities of these reaction models, developed by the CLAS Collaboration, for the credible 
extraction of the $\gamma_vpN^\ast$ electrocouplings from independent studies of different electroproduction channels.

\subsection{Insights into the Dressed-Quark Mass Function from the $\gamma_vpN^\ast$ Electrocouplings}
\label{ehm_resel}

\begin{figure}[t]
\includegraphics[width=0.85\textwidth, left]{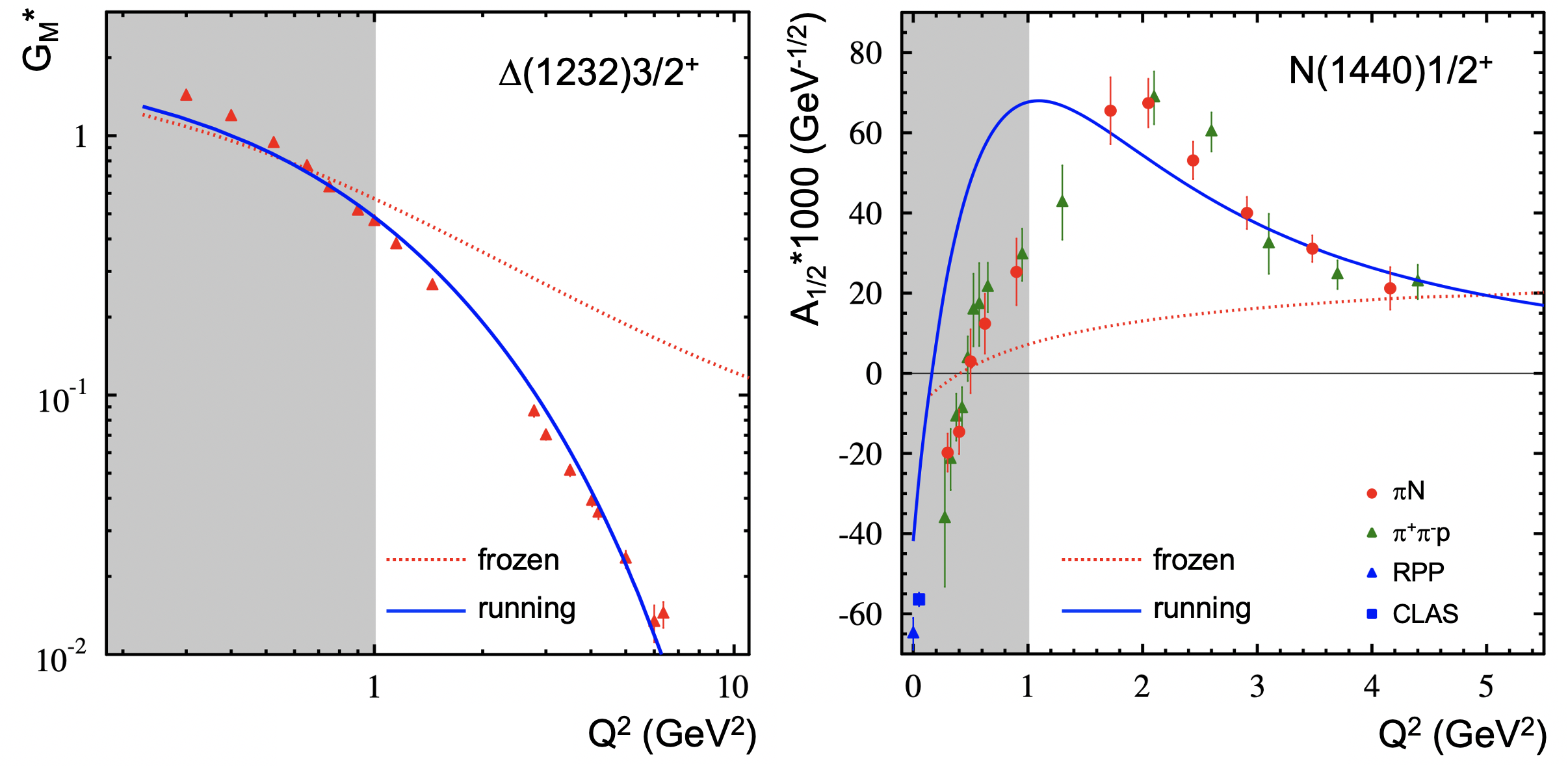}
\caption{Description of the results for the $N \to \Delta$ magnetic transition form factor G$^\ast_{\text{M}}$ (left) and the electrocoupling 
amplitude $A_{1/2}$ for the $N\to N(1440)1/2^+$ (right) achieved using CSMs~\cite{Wilson:2011aa, Segovia:2013uga, Segovia:2014aza, Segovia:2015hra}. 
Results obtained with a momentum-independent (frozen) dressed-quark mass~\cite{Wilson:2011aa, Segovia:2013uga} (dotted red curves) are compared with 
QCD-kindred results (solid blue curves) obtained with the momentum-dependent quark mass function in Figure~\ref{alpha_mass_running}. The
electrocoupling data were taken from References~\cite{CLAS:2009ces,Villano:2009sn,CLAS:2014fml} -- $\pi N$ electroproduction; and 
References~\cite{Mokeev:2015lda, CLAS:2012wxw, VictorMokeevPresentation1, VictorMokeevPresentation2} -- $\pi^+\pi^-p$ electroproduction. 
The photocouplings for the $N(1440)1/2^+$ are from the RPP~\cite{Workman:2022ynf} and from Reference~\cite{Dugger:2009pn} -- blue square and triangle,
respectively. The ranges of $Q^2$ where the contributions from the meson-baryon cloud remain substantial are highlighted in gray.}
\label{csm_delta_roper}
\end{figure}

Results on the $Q^2$ evolution of the $\gamma_vpN^\ast$ electrocouplings available from experiments performed during the 6-GeV era at JLab have 
already had a substantial impact on understanding the sQCD dynamics responsible for the saturation of the running coupling $\alpha_s$, $N^\ast$
structure, and the generation of a significant portion of hadron mass~\cite{Proceedings:2020fyd, Burkert:2019bhp}. Analyses of these results have
revealed $N^\ast$ structure to emerge from a complex interplay between the inner core of three dressed quarks and an outer meson-baryon cloud
\cite{Proceedings:2020fyd, Burkert:2019bhp, Aznauryan:2014xea, Mokeev:2015lda}. Successful descriptions of the data on the dominant 
$N \to \Delta(1232)3/2^+$ magnetic transition form factor~\cite{Aznauryan:2009mx, Villano:2009sn} and the electrocouplings of the $N(1440)1/2^+$
\cite{Aznauryan:2009mx, CLAS:2014fml, Mokeev:2020vab} have been achieved using CSMs~\cite{Wilson:2011aa, Segovia:2013uga, Segovia:2014aza,
Segovia:2015hra} for $Q^2 > 1.0$~GeV$^2$ and $Q^2 > 2.0$~GeV$^2$, respectively (see Figure~\ref{csm_delta_roper}). These $Q^2$ ranges correspond 
to the distance scales where contributions from the quark core to the resonance structure come to dominate. Since the CSM evaluations account 
for the contributions from only the quark core, they can only reasonably be confronted with experimental results in the higher-$Q^2$ range, where 
the quark core contributions to $N^\ast$ structure dominate over those from the meson-baryon cloud.

The sensitivity of the electroexcitation amplitudes to the momentum-dependence of the quark mass function is dramatically illustrated by 
Figure~\ref{csm_delta_roper}, which deliberately shows results obtained with $M(k)=\,$constant~\cite{Wilson:2011aa, Segovia:2013uga} and $M(k)$ 
from Figure~\ref{alpha_mass_running}~\cite{Segovia:2014aza, Segovia:2015hra}. The $M(k)=0.36\,$GeV results were computed first in order to provide 
a ``straw-man'' benchmark against which the subsequent realistic $M(k)$ results could be contrasted. The constant-mass results (dotted red curves 
in Figure~\ref{csm_delta_roper}) overestimate the data on the $N \to \Delta$ magnetic transition form factor for $Q^2 \gtrsim 1\,$GeV$^2$. The
discrepancy increases with $Q^2$, approaching an order of magnitude difference in the ratio at 5\,GeV$^2$. Moreover, whilst reproducing the zero 
in $A_{1/2}$ for the $N(1440)1/2^+$, the frozen mass result is otherwise incompatible with the data. Plainly, therefore, the data speak against 
dressed-quarks with a frozen mass. On the other hand, in both cases, the transition form factors are well described by an internally consistent 
CSM calculation built upon the mass function in Figure~\ref{alpha_mass_running} -- see the solid blue curves in Figure~\ref{csm_delta_roper}. These
observations confirm the statements made above, \textit{viz}.\ nucleon resonance electroexcitation amplitudes are keenly sensitive to the form of 
the running quark mass. Moreover, the agreement with the larger-$Q^2$ data clearly points to a dominance of the dressed-quark core of the nucleon 
resonances in the associated domains.

It is worth stressing that the CSM results for the $\Delta(1232)3/2^+$ and $N(1440)1/2^+$ electroexcitation amplitudes were obtained using the same
dressed-quark mass function, \textit{i.e}., $M(k)$ in Figure~\ref{alpha_mass_running}: indeed, the theory analyses of both transitions used precisely
the same framework. The common quark mass function matches that obtained by solving the quark gap equation in Figure~\ref{gap_equations} with a 
kernel built from the best available inputs for~\cite{Binosi:2016wcx}: the gluon two-point function, running coupling, and dressed gluon-quark vertex.
Moreover, the same mass function was also used in the successful description of the experimental results on nucleon elastic electromagnetic form 
factors~\cite{Segovia:2014aza, Cui:2020rmu}, and axial and pseudoscalar form factors~\cite{Chen:2021guo, Chen:2022odn}. Such a mass function is also 
a key element in an \textit{ab initio} treatment of pion electromagnetic elastic and transition form factors~\cite{Chang:2013nia, Chang:2013pq,
Raya:2015gva}.

These CSM results for meson and baryon properties, both ground and excited states, are part of a large body of mutually consistent predictions.
Their success in describing and explaining data relating to such a diverse array of systems provides strong evidence in support of the position 
that dressed-quarks, with dynamically generated running masses, are the appropriate degrees-of-freedom for use in the description of the mass and
structure of all hadrons. This realization is one of the most important achievements of hadron physics during the past decade, and it was only 
accomplished through numerous synergistic interactions between experiment, phenomenology, and theory. 

\subsection{Novel Tests of CSM Predictions}

In 2019, CSM predictions became available for the electrocouplings of the $\Delta(1600)3/2^+$ \cite{Lu:2019bjs}. This baryon may be interpreted in 
quantum field theory as a state with aspects of the character of a first radial excitation of the $\Delta(1232)3/2^+$~\cite{Qin:2018dqp, Liu:2022ndb},
for which CSM electrocoupling results became available earlier~\cite{Segovia:2014aza} and are discussed above.

\begin{figure}[t]
\centering
\includegraphics[width=1.0\textwidth, left]{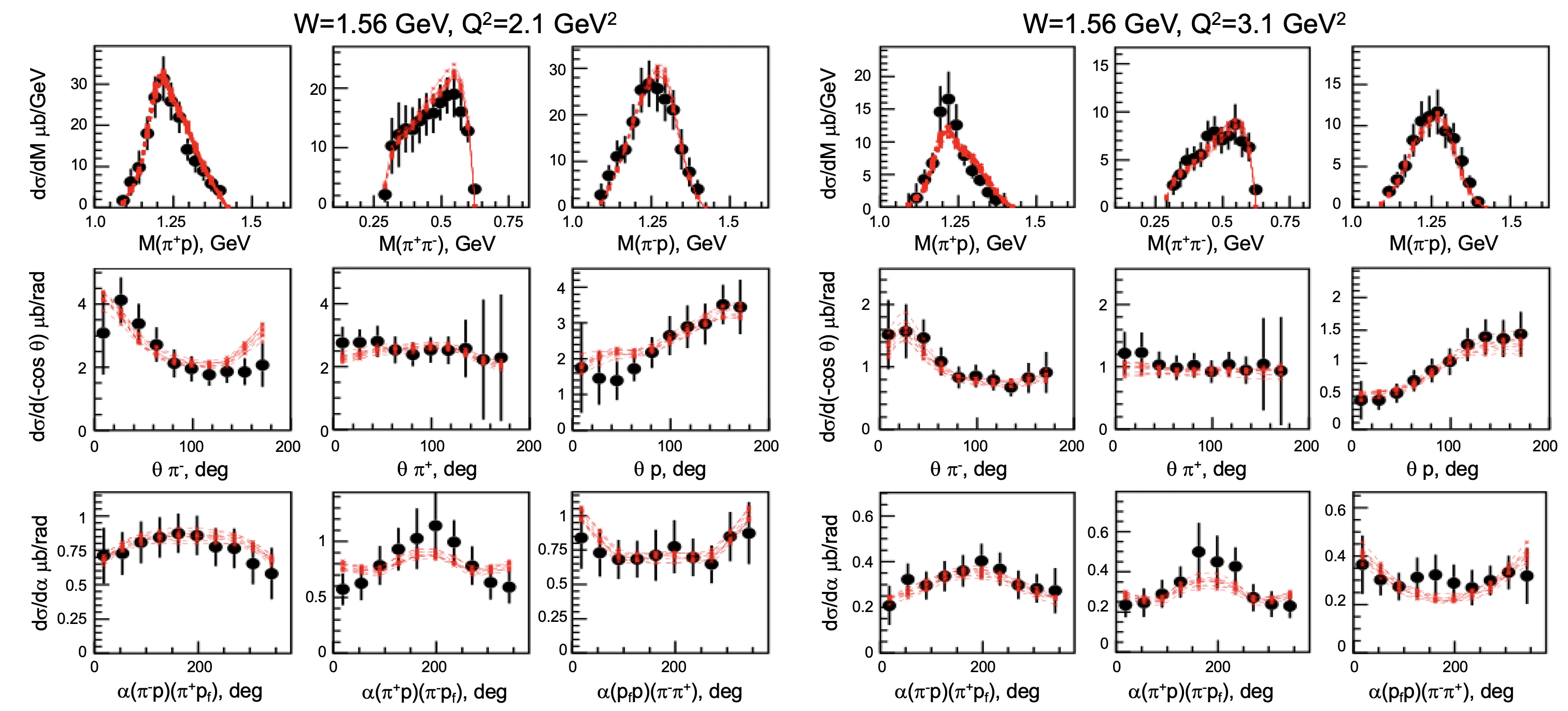}
\vspace{-4mm}
\caption{Regarding extraction of $\Delta(1600)3/2^+$ electrocouplings, representative examples of the nine independent one-fold differential cross
sections available from the $\pi^+\pi^-p$ measurements with CLAS~\cite{CLAS:2017fja,Trivedi:2018rgo} at two different $Q^2$ values, along with the 
data fits within the data-driven meson-baryon JM reaction model~\cite{CLAS:2012wxw, Mokeev:2008iw, Mokeev:2020vab}.
\label{pippimp_data}}
\end{figure}

No relevant experimental results were available when the $\Delta(1600)3/2^+$ predictions were made. The first (and still preliminary) results for the
$\Delta(1600)3/2^+$ electrocouplings only became available in the first half of 2022~\cite{VictorMokeevPresentation, VictorMokeevPresentation1,
VictorMokeevPresentation2}. They were extracted from the analysis of $\pi^+\pi^-p$ electroproduction off protons measured with the CLAS detector 
\cite{CLAS:2017fja, Trivedi:2018rgo} for $W$ from 1.4--2.1\,GeV and $Q^2$ from 2--5\,GeV$^2$. Nine independent one-fold differential cross 
sections were analyzed in each ($W$,$Q^2$) bin. The final-state hadron kinematics is fully determined by the five-fold differential cross sections. 
The one-fold differential cross sections were obtained by integrating the five-fold differential cross sections over different sets of four
kinematic variables~\cite{CLAS:2012wxw, Mokeev:2008iw}. For extraction of the electrocouplings, it was necessary to fit the data for the three 
invariant mass distributions for the different pairs of final-state hadrons, the distributions of the final-state hadrons over the CM polar angles
$\theta_i$ ($i=\pi^+,\pi^-,p_f$), and the distributions over the three CM angles $\alpha_{[i],[j]}$ between the two planes: one of which [$i$] is 
the reaction plane defined by the three-momentum of the $\gamma_v$ and one of the final state hadrons, and the second [$j$] is determined by the 
three-momenta of the other two final-state hadrons for the three possible choices of the hadron pairs. Representative examples of the data measured 
are shown in Figure~\ref{pippimp_data} at the $W$-bins closest to the Breit-Wigner mass of the $\Delta(1600)3/2^+$ and in different bins of $Q^2$.

The $N^\ast$ electrocouplings on the domain $W < 1.65$\,GeV were obtained from the fit of the differential $\pi^+\pi^-p$ photo- and electroproduction
cross sections carried out within the framework of the data-driven JM meson-baryon reaction model~\cite{Mokeev:2015lda, CLAS:2012wxw, Mokeev:2008iw,
Burkert:2007kn, Ripani:2000va}. This model has been developed by the CLAS Collaboration for the extraction of nucleon resonance electrocouplings and 
their partial hadronic decay widths to the $\pi\Delta$ and $\rho p$ final states. Within the JM model, the full 3-body $\pi^+\pi^-p$ 
electroproduction amplitude includes the contributions from $\pi^-\Delta^{++}$, $\rho p$, $\pi^+\Delta^0$, $\pi^+N(1520)3/2^-$, and
$\pi^+N(1685)5/2^+$, with subsequent decays of the unstable intermediate hadrons. It also contains direct $2\pi$ photo-/electroproduction processes,
where the final $\pi^+\pi^-p$ state is created without the generation of unstable intermediate hadrons. Here the nucleon resonances contribute to the
$\pi^-\Delta^{++}$, $\pi^+\Delta^0$, and $\rho p$ channels.

Modeling of the non-resonant contributions is described in References~\cite{Mokeev:2008iw, CLAS:2012wxw, Burkert:2007kn, Ripani:2000va}. For the
resonant contributions, the JM model includes all four-star Particle Data Group (PDG) $N^\ast$ states with observed decays to $\pi\pi N$, as well 
as the new $N^\prime(1720)3/2^+$ resonance~\cite{Mokeev:2020hhu, Mokeev:2021dab} observed in the combined analysis of $\pi^+\pi^-p$ photo- and
electroproduction data. The resonant amplitudes are described within the unitarized Breit-Wigner \textit{ansatz}~\cite{CLAS:2012wxw}, thereby 
ensuring consistency with restrictions imposed by the general unitarity condition. The JM model offers a good description of the $\pi^+\pi^-p$
differential cross sections in the entire kinematic area covered by the data at $W < 2.1$\,GeV and $Q^2 < 5$\,GeV$^2$. All of the electrocouplings 
extracted from the $\pi^+\pi^-p$ data (published, in part, also in the PDG) have been determined using the JM reaction model.

In the analyses of the $\pi^+\pi^-p$ data~\cite{CLAS:2017fja, Trivedi:2018rgo}, the following quantities were varied: the $\gamma_vpN^\ast$ 
electrocouplings for the resonances in the mass range $< 1.75\,$GeV, their partial hadronic decay widths into the $\pi\Delta$ and $\rho p$ final 
states, their total decay widths, and the non-resonant parameters of the JM model. For each trial attempt at a data description, $\chi^2/d.p.$ 
($d.p.$ = data point) was computed by the comparison between the measured and computed nine one-fold differential cross sections. In the fits, 
the computed cross sections closest to the data were selected by requiring $\chi^2/d.p.$ to be below a predetermined threshold, ensuring that 
the spread of the selected phenomenological fit cross sections lies within the data uncertainties for most experimental data points.
Representative examples are shown by the family of curves in Figure~\ref{pippimp_data}.

\begin{figure}[t]
\centering
\includegraphics[width=0.95\textwidth, left]{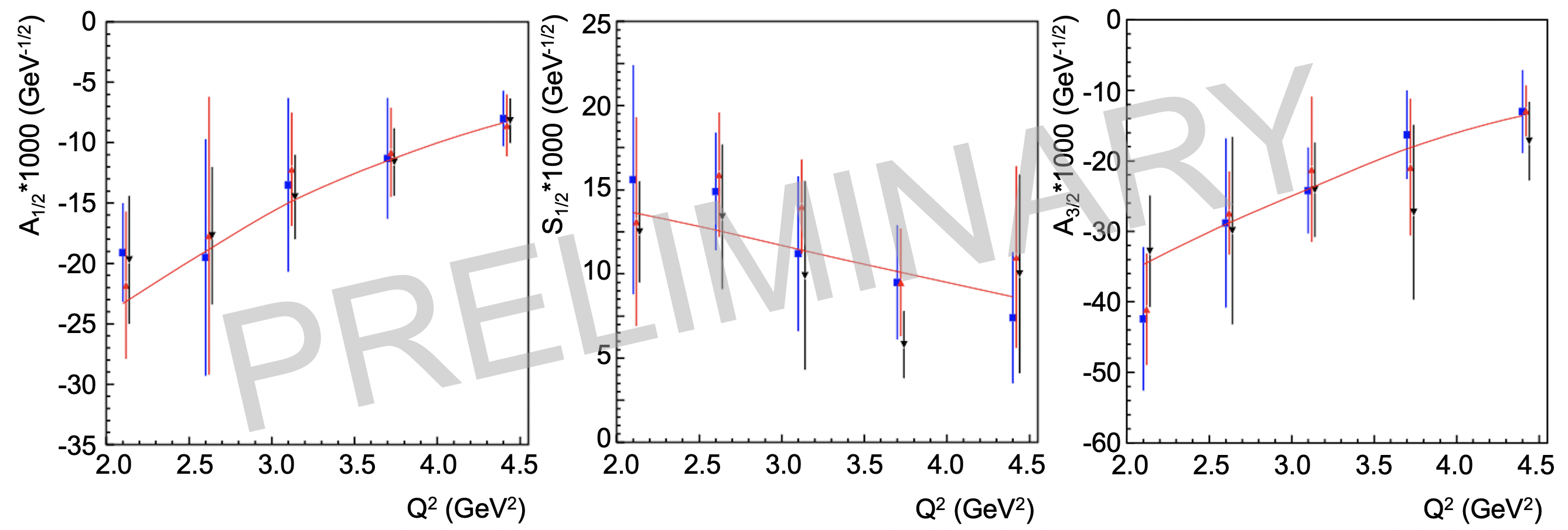}
\caption{$\Delta(1600)3/2^+$ electrocouplings with their assigned uncertainties, determined from independent analysis of the $\pi^+\pi^-p$ 
differential cross sections in three overlapping $W$ intervals: 1.46--1.56\,GeV (filled blue squares), 1.51--1.61\,GeV (filled red triangles), and 
1.56--1.66\,GeV (filled black triangles)~\cite{VictorMokeevPresentation, VictorMokeevPresentation1, VictorMokeevPresentation2}. 
CSM predictions \cite{Lu:2019bjs} are drawn as solid red curves.
\label{delta1600_electrocouplings}}
\end{figure}

The electrocouplings for the computed cross sections selected from the data fits were averaged together and their means were treated as the
experimental value. The RMS width of the determined electrocouplings was assigned as the corresponding uncertainty. The preliminary results of 
this extraction for the $\Delta(1600)3/2^+$~\cite{VictorMokeevPresentation, VictorMokeevPresentation1, VictorMokeevPresentation2} are shown in 
Figure~\ref{delta1600_electrocouplings}, wherein they are compared with the CSM predictions obtained three years earlier~\cite{Lu:2019bjs}. 
These results were determined for overlapping $W$-intervals: 1.46--1.56\,GeV, 1.51--1.61\,GeV, and 1.56--1.66\,GeV for $Q^2$ from 2--5\,GeV$^2$. 
The non-resonant contributions in these $W$-intervals are different. The electrocouplings determined from the independent fits of the data within 
the three $W$-intervals are consistent, establishing their reliability and confirming the CSM predictions. This success has markedly strengthened 
the body of evidence that indicates that detailed information can be obtained on the momentum dependence of the dressed-quark mass function from 
sound data on $N^\ast$ electroproduction, and, therefrom, deep insights into the character of EHM in the SM. 

\section{Studies of $N^\ast$ Structure in Experiments with CLAS12 and Beyond}
\label{prospects}

\begin{figure}[t]
\includegraphics[width=0.75\textwidth, left]{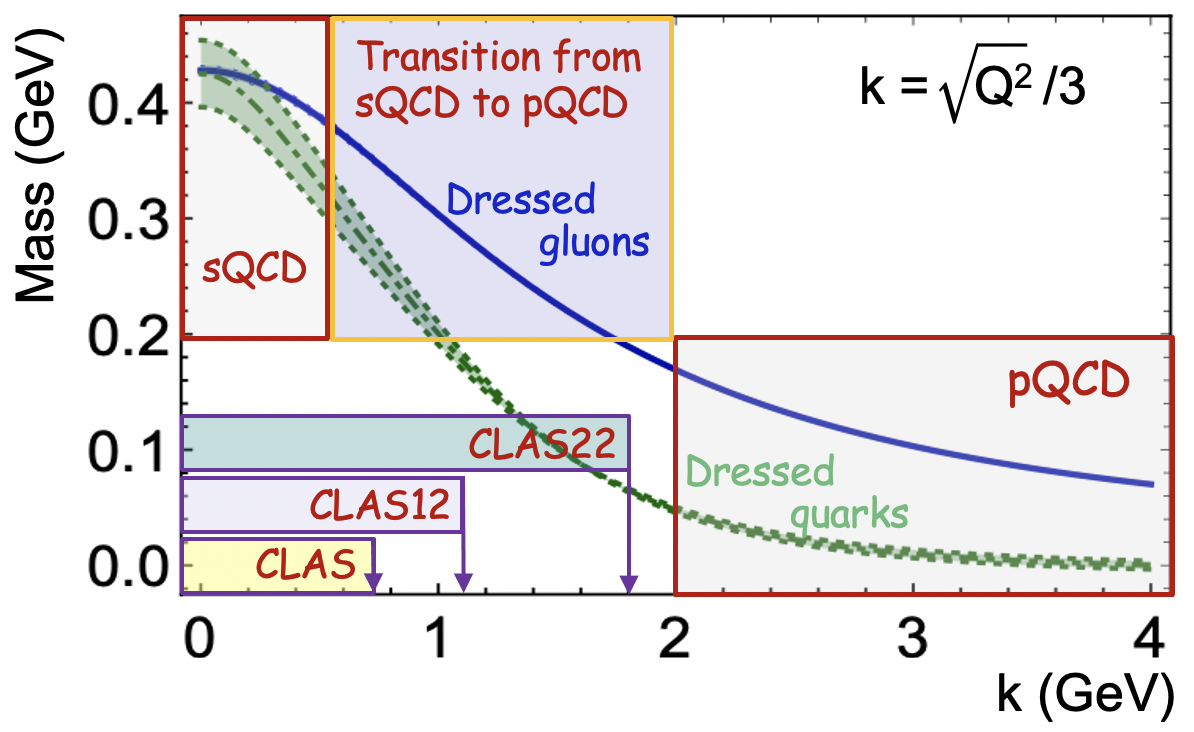}
\includegraphics[width=0.75\textwidth, left]{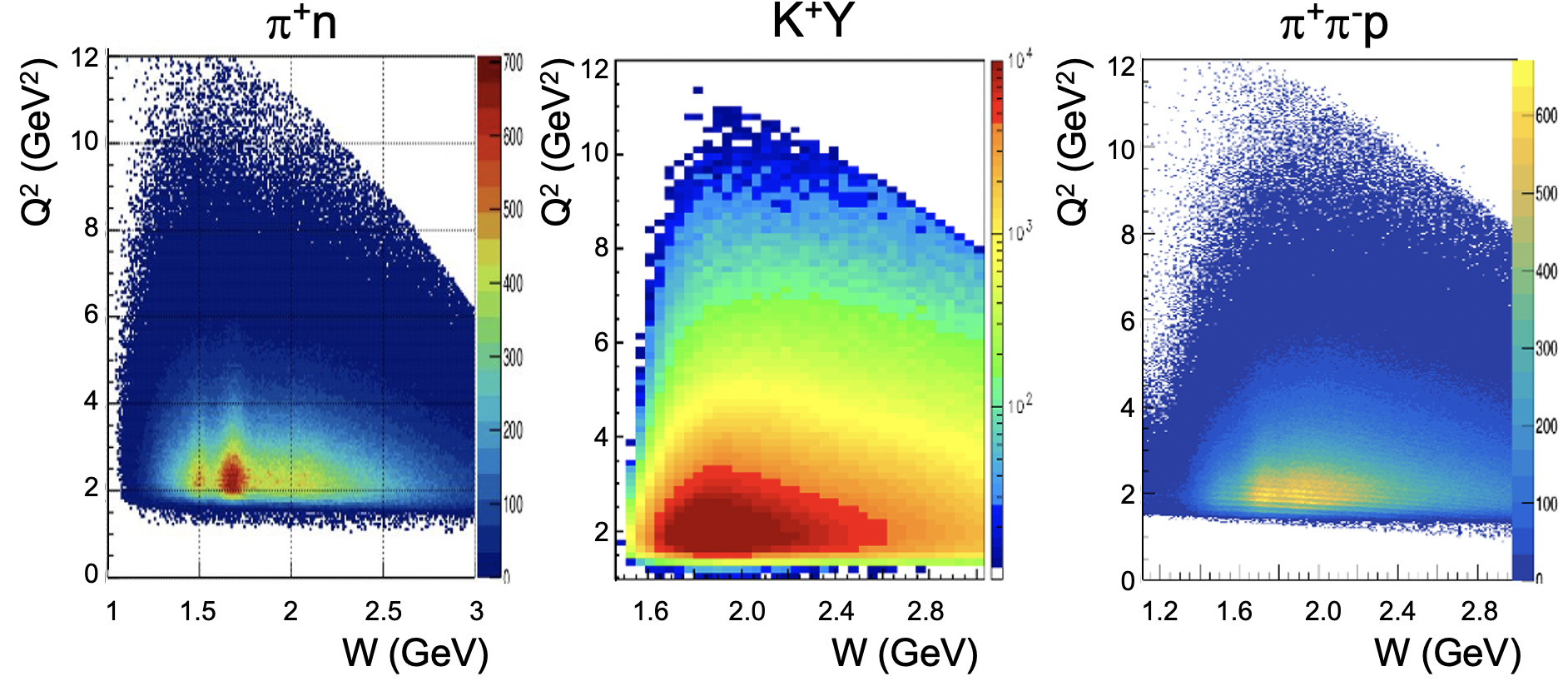}
\caption{(Top) Momentum ranges accessible in the exploration of the momentum dependence of the dressed quark mass function using results on the 
$Q^2$ evolution of $\gamma_vpN^\ast$ electrocouplings. The range of $k$ covered by available data is mostly from experiments with CLAS, shown in
yellow. The expected reach of CLAS12 experiments is shown in purple and that achievable after a proposed increase of the JLab beam energy to 22~GeV 
in cyan. (Bottom) Yields of representative exclusive meson electroproduction channels available from the experiments with the CLAS12 detector.}
\label{running_mass_reach}
\end{figure}

Most results on the $N^\ast$ electrocouplings have been obtained for $Q^2 < 5$\,GeV$^2$. Detailed comparison of these results with the CSM 
predictions allow for exploration of the quark mass function within the range of quark momenta $< 0.75$\,GeV, assuming equal sharing of the 
virtual-photon momentum-transfer between the three dressed quarks in the transition between the ground and excited nucleon states. The results on 
the resonance electrocouplings in this range of quark momentum, shown in Figure~\ref{running_mass_reach} (top), cover distances over which less 
than 30\% of hadron mass is generated~\cite{Roberts:2021xnz, Roberts:2020hiw}.

In the northern spring of 2018, after completion of the 12-GeV-upgrade project, measurements with the CLAS12 detector in Hall~B at JLab commenced 
\cite{Mokeev:2022xfo, Carman:2020qmb, Burkert:2020akg}. Currently, CLAS12 is the only facility in the world capable of exploring exclusive meson
electroproduction in the resonance region, exploiting the highest $Q^2$ ever achieved for these processes. Ongoing experiments with electron beam 
energies up to 11\,GeV with CLAS12 offer a unique opportunity to obtain information on the electrocouplings of the most prominent $N^\ast$ states 
in the mass range up to 2.5\,GeV at $Q^2$ up to 10\,GeV$^2$ from the exclusive $\pi N$, $KY$ ($Y$=$\Lambda$ or $\Sigma$), $K^\ast Y$, $KY^\ast$, 
and $\pi^+\pi^-p$ channels~\cite{CLAS:12a,CLAS:12b,CLAS:12c,Carman:2016hlp,Carman:2018fsn}. $Q^2$ versus $W$ event distributions for these exclusive
reaction channels measured with CLAS12 at a beam energy of $\sim$11\,GeV are shown in Figure~\ref{running_mass_reach} (bottom). The first 
results from the CLAS12 $N^\ast$ program (at lower beam energies of 6.5/7.5\,GeV) have recently been published on the beam-recoil hyperon 
transferred polarization in $K^+Y$ electroproduction~\cite{CLAS:2022yzd}. The increase in the $Q^2$-coverage for results on the electrocouplings
from CLAS12 will enable exploration of the dressed quark mass within the range of quark momenta where roughly 50\% of hadron mass is expected to 
be generated (see Figure \ref{running_mass_reach} (top)).

\begin{figure}[t]
\centering
\includegraphics[width=0.65\textwidth, left]{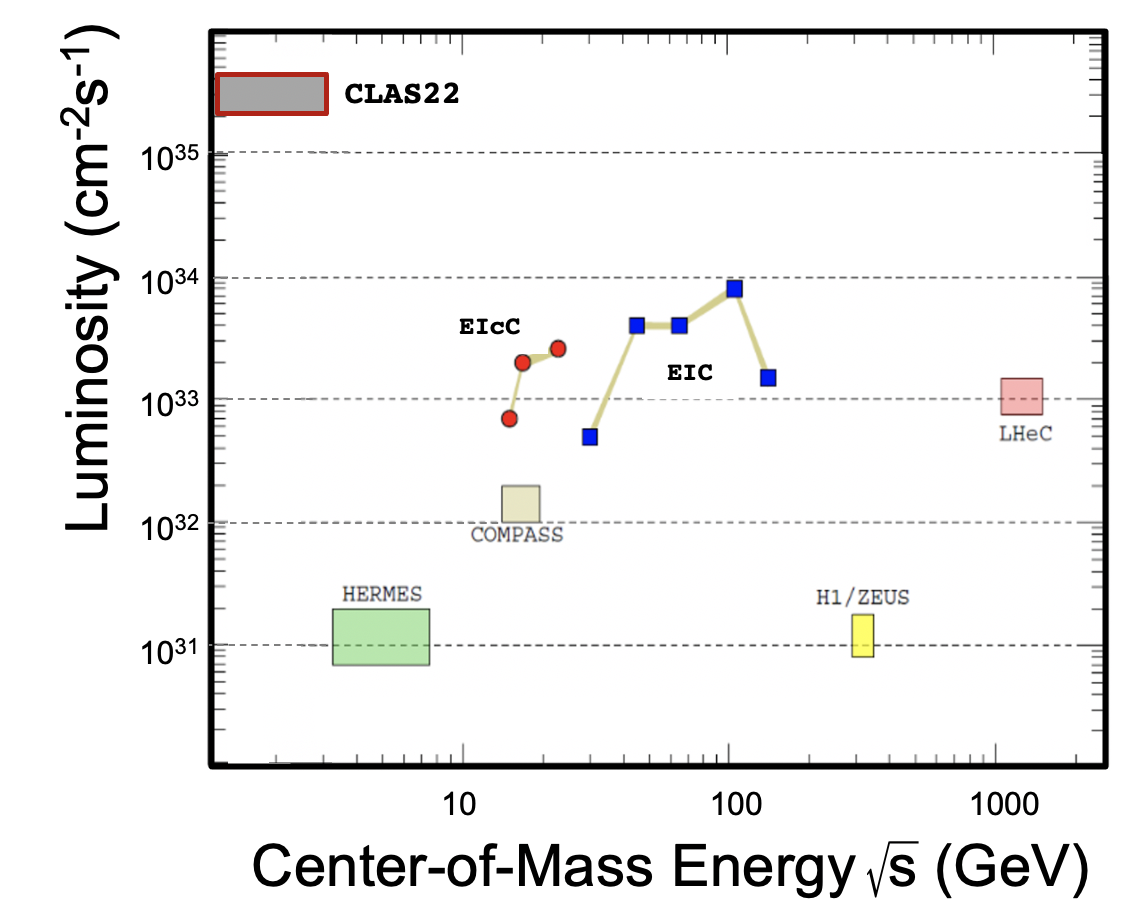}
\caption{Luminosity versus CM energy in lepton-proton collisions for existing and foreseeable facilities capable of exploring hadron structure 
in measurements with large-acceptance detectors.
\label{facilities}}
\end{figure}

In order to solve the challenging SM problems relating to EHM, the dressed quark mass function should be charted over the entire quark momentum
range up to $\approx2\,$GeV. This is the domain of transition from strong to perturbative QCD (see Figure~\ref{running_mass_reach} (top)) and 
where dressed quarks and gluons become the relevant degrees-of-freedom as $\alpha_s/\pi \to 1$, approaching the sQCD saturation regime (see 
Figure~\ref{alpha_mass_running}). This objective requires a further extension of the electrocoupling measurements up to $Q^2 \approx 30$\,GeV$^2$. 
Discussions and planning are currently underway, focusing on an energy-increase of the JLab accelerator to a beam energy of 22\,GeV, after the 
completion of the experiments planned for the 12-GeV program. Initial simulations of $\pi N$, $KY$, and $\pi^+\pi^-p$ electroproduction at 22\,GeV
using the existing CLAS12 detector at a luminosity up to $(2-5) \times 10^{35}$~cm$^{-2}$s$^{-1}$ have shown that a measurement program of 
1-2~years duration would enable measurements of sufficient statistical accuracy to determine the $\gamma_vpN^\ast$ electrocouplings of the most
prominent $N^\ast$ states over this full kinematic range. 

Figure~\ref{facilities} shows the luminosity versus CM energy in lepton-proton collisions for existing and foreseeable facilities capable of
exploring hadron structure in measurements with large-acceptance detectors. The luminosity requirements for the extraction of electrocouplings 
within the $Q^2$ range 10-30\,GeV$^2$ exceed by more than an order-of-magnitude the maximum luminosity planned for experiments with the EIC
\cite{AbdulKhalek:2021gbh} and EicC~\cite{Anderle:2021wcy} $ep$ colliders and even more for other facilities. The combination of a high duty-factor
JLab electron beam at 22\,GeV with the capacity to measure exclusive electroproduction reactions at luminosities of 
$(2-5) \times 10^{35}$~cm$^{-2}$s$^{-1}$ using a large-acceptance detector, would make a 22\,GeV JLab unique. It would be the only facility in the
world able to explore the evolution of hadron structure over essentially the full range of distances where the transition from strong-coupling QCD 
to the weak-field domain is expected to occur.

The increase of the JLab energy to 22\,GeV, pushing the current CLAS12 detector capabilities to measure exclusive electroproduction to the highest 
possible luminosity and extending the available reaction models used for the extraction of the electrocouplings, will offer the only foreseeable
opportunity to explore how the dominant part of hadron mass (up to 85\%) and $N^\ast$ structure emerge from QCD. This would make an 
energy-upgraded JLab at 22~GeV the ultimate QCD-facility at the luminosity frontier. 

\section{Conclusions and Outlook}

Baryons are the most fundamental three-body systems in Nature. If we do not understand how QCD generates these bound states of three dressed 
quarks, then our understanding of Nature is incomplete. Remarkable progress has been achieved in recent decades through the studies of the 
structure of the ground and excited nucleon states in experiments at JLab during the 6-GeV era~\cite{Burkert:2019opk, Burkert:2018oyl, 
Burkert:2019bhp, Mokeev:2022xfo, Carman:2020qmb, Aznauryan:2011ub, Aznauryan:2011qj, Proceedings:2020fyd}. These experiments have provided a 
large array of new opportunities for QCD-connected hadron structure theory by opening a door to the exploration of many hitherto unseen facets of 
the strong interaction in the regime of large running coupling, \textit{i.e}., $\alpha_s/\pi \gtrsim 0.2$, by providing the results on the 
$\gamma_vpN^\ast$ electrocouplings for numerous $N^\ast$ states, with different quantum numbers and structural features. 

High-quality meson electroproduction data from the 6-GeV era at JLab have enabled determination of the electrocouplings of most nucleon resonances 
in the mass range up to 1.8\,GeV for $Q^2 < 5\,$GeV$^2$ (up to 7.5\,GeV$^2$ for the $\Delta(1232)3/2^+$ and $N(1535)1/2^-$). Consistent results on 
the $Q^2$ evolution of these electrocouplings from analyses of $\pi^+n$, $\pi^0 p$, $\eta p$, and $\pi^+\pi^-p$ electroproduction have demonstrated 
the capability of the reaction models employed to extract the electrocouplings in independent studies of all of these different exclusive channels. 
Above, we have sketched how comparisons between the experimental results on the $Q^2$ evolution of the $\gamma_vpN^\ast$ electroexcitation 
amplitudes and QCD-connected theory have vastly improved our understanding of the momentum dependence of the dressed quark mass function, which is 
one of the three pillars of EHM. The remaining two pillars are the running gluon mass and the QCD effective charge, and these entities, too, are 
constrained by the electroexcitation data. 

A good description of the $\Delta(1232)3/2^+$ and $N(1440)1/2^+$ electrocouplings has been achieved using CSMs in the full range of photon
virtualities where the structure of these excited states is principally determined by contributions from a core of three dressed quarks. The 
successful description of the electrocouplings for nucleon resonances of different structure, spin+isospin flip for the $\Delta(1232)3/2^+$ and the
first radial excitation of three dressed quarks for the $N(1440)1/2^+$, was achieved with the same dressed quark mass function. This mass function 
is determined by QCD dynamics, and such a running mass has also been used in the successful description of data on elastic electromagnetic nucleon 
and pion form factors, as well as for the description of the nucleon axial form factor $G_A$. In thereby arriving at a unification of diverse
observables, one obtains compelling evidence in support of the momentum dependence of the dressed quark mass used to describe the results on the 
$Q^2$ evolution of the electrocouplings. 

This impressive hadron physics achievement in the past decade was accomplished through synergistic efforts between experiment, phenomenology, and 
QCD-connected hadron structure theory. In 2019 CSMs provided parameter-free predictions for the electrocouplings of the $\Delta(1600)3/2^+$. 
There were no experimental results available at that time. The first, preliminary results on the $\Delta(1600)3/2^+$ electrocouplings extracted 
from the data on $\pi^+\pi^-p$ electroproduction are reported herein. They have strikingly confirmed the CSM predictions.    

Most results for the $\gamma_vpN^\ast$ electrocouplings are currently available for $Q^2 < 5$~GeV$^2$, allowing for the exploration of the dressed
quark mass within the limited range of quark momenta where less than 30\% of hadron mass is expected to be generated. Experiments on exclusive 
meson electroproduction in the resonance region are now in progress with the CLAS12 detector in Hall~B at JLab, following completion of the 
12-GeV-upgrade project. CLAS12 is the only facility in the world capable of obtaining the electrocouplings of all prominent $N^\ast$ states in the
still unexplored $Q^2$ range from 5--10\,GeV$^2$ from measurements of $\pi N$, $\eta p$, $\pi^+\pi^-p$, and $KY$ electroproduction. These data 
will probe the dressed quark mass function at quark momenta up to $\approx 1.1\,$GeV, a domain where up to 50\% of hadron mass is generated.
  
In order to solve the problem of EHM, a key challenge within the SM, the dressed quark mass function should be mapped over the entire range of 
quark momenta up to $\approx$2~GeV, where the transition from strong to perturbative QCD takes place and where gluon and quark quasiparticles 
with dynamically generated running masses emerge as $\alpha_s/\pi \to$ 1. This requires an extension of existing and anticipated data so that it 
covers the $Q^2$-domain from 10--30~GeV$^2$. Explorations of the possibility to increase the JLab beam energy to 22\,GeV are now in progress. 
Such a machine would enable coverage over the desired $Q^2$ range within the region of $W < 2.5$~GeV. Simulations with the existing CLAS12 
detector configuration for the exclusive $\pi N$, $KY$, and $\pi^+\pi^-p$ electroproduction channels at 22\,GeV beam energy and a luminosity of 
$(2-5) \times 10^{35}$~cm$^{-2}$s$^{-1}$ show that, with beam-times of 1-2~years, differential cross section and polarization asymmetry measurements 
of sufficient statistical precision can be achieved to extract electrocouplings of all prominent resonances up to 30\,GeV$^2$. Both the EIC and EicC
$ep$ colliders would need much higher and foreseeably unreachable luminosities than currently envisaged in order to carry out such a program. 
The combination of a high duty-factor 22\,GeV JLab electron beam and the capability to measure exclusive electroproduction events at high 
luminosities with a large-acceptance detector would make JLab the ultimate QCD-facility at the luminosity frontier. It would be unique in 
possessing the capacity to explore the evolution of hadron mass and structure over the full range of distances where the transition from sQCD to 
pQCD is expected. 

Drawing a detailed map of proton structure is important because the proton is Nature’s only absolutely stable bound state. However, understanding 
how QCD's simplicity explains the emergence of hadron mass and structure requires investment in a facility that can deliver precision data on much 
more than one of Nature’s hadrons. An energy-upgraded JLab complex is the only envisaged facility that could enable science to produce a sufficient
quantity of precise structure data on a wide range of hadrons with distinctly different quantum numbers and thereby move into a new realm of 
understanding. There is elegance in simplicity and beauty in diversity. If QCD possesses both, then it presents a very plausible archetype for 
taking science beyond the Standard Model. In that case, nuclear physics at JLab\,20$+$ has the potential to deliver an answer that takes science 
far beyond its current boundaries.



\vspace{6pt}




\funding{Work supported by:
U.S. Department of Energy (contract no. DE-AC05-06OR23177) (DSC, VIM),
National Science Foundation (NSF grant PHY\,10011349) (RWG),
and
National Natural Science Foundation of China (grant no.\,12135007) (CDR).
}

\acknowledgments{

This contribution is based on results obtained and insights developed through collaborations with many people, to all of whom we are greatly indebted.
%
%
}

\conflictsofinterest{The authors declare no conflict of interest.}



\abbreviations{Abbreviations}{
The following abbreviations are used in this manuscript:\\

\noindent
\begin{tabular}{@{}ll}
CM & center-of-mass \\
CSM & continuum Schwinger function method \\
DCSB & dynamical chiral symmetry breaking \\
{$d.p.$} & data point \\
EHM & emergence of hadron mass \\
EIC & Electron-Ion Collider (at Brookhaven National Laboratory) \\
EicC & Electron-ion collider China \\
HB & Higgs boson \\
JLab & Thomas Jefferson National Accelerator Facility (Jefferson Laboratory)\\
JM & JLab-Moscow State University \\
lQCD & lattice-regularized quantum chromodynamics \\
NG (mode/boson) & Nambu-Goldstone (mode/boson) \\
PDFs & Particle Distribution Functions \\
PDG & Particle Data Group (and associated publications) \\
pQCD & perturbative QCD \\
QCD & quantum chromodynamics \\
RMS & root mean square \\
RPP & Review of Particle Properties (and associated publications) \\
sQCD & strong QCD \\
SM & Standard Model of particle physics
\end{tabular}
}

\begin{adjustwidth}{-\extralength}{0cm}
\reftitle{References}


\bibliography{CollectedBiB1}

\end{adjustwidth}
\end{document}